\documentclass{aa}

\usepackage[dvipsnames]{xcolor}
\usepackage{color}

\usepackage{natbib}
\usepackage{graphicx}
\usepackage{txfonts}
\usepackage{amsmath}
\usepackage{chemmacros}
\usepackage{lscape}
\usepackage{ulem}

\usepackage{hyperref}

\hypersetup{
    colorlinks=true,
    linkcolor=MidnightBlue,
    filecolor=MidnightBlue,      
    urlcolor=MidnightBlue,
    citecolor=MidnightBlue
}

\colorlet{myred}{red!60!gray}
\colorlet{mygreen}{green!60!gray}

\newcommand{\msun}{M$_\odot$}
\def\kms {km~s$^{-1}$}
\newcommand*{\ditto}{---\texttt{"}---}
\newcommand{\e}[1]{\times 10^{#1}}

\begin{document}

   \title{Carbon monoxide formation and cooling in supernovae}

   \author{S. Liljegren\inst{1,2}
   \and
   A. Jerkstrand\inst{1,3}
   \and
   J. Grumer\inst{2}}

   \institute{The Oskar Klein Centre, Department of Astronomy, Stockholm University, Albanova 10691, Stockholm, Sweden \label{inst1}
   \\ \email{sofie.liljegren@astro.su.se}
   \and 
    Theoretical Astrophysics, Department of Physics and Astronomy, Uppsala University, Box 516, SE-751 20 Uppsala, Sweden \label{inst2}
        \and  
        Max Planck Institute for Astrophysics, Karl-Schwarzschild-Straße 1, 85748 Garching, Germany \label{inst3}}

   \date{Received; accepted }

% \abstract{}{}{}{}{} 
% 5 {} token are mandatory
 
  \abstract
   {The inclusion of molecular physics is an important  piece that tends to be missing from the puzzle when modeling the spectra of supernovae (SNe). Molecules have both a direct impact on the spectra, particularly in the infrared, and an indirect one as a result of their influence on certain physical conditions, such as temperature.}
   { 
   In this paper, we aim to investigate molecular formation and non-local thermodynamic equilibrium (NLTE) cooling, with a particular focus on CO, the most commonly detected molecule in supernovae. We also aim to determine the dependency of supernova chemistry on physical parameters and the relative sensitivity to rate uncertainties.
   }
   {
   We implemented a chemical kinetic description of the destruction and formation of molecules into the SN spectral synthesis code \textsc{Sumo}. In addition, selected molecules were coupled into the full NLTE level population framework and, thus, we incorporated molecular NLTE cooling into the temperature equation. 
    We produced a test model of the CO formation in SN 1987A between 150 and 600 days and investigated the sensitivity of the resulting molecular masses to the input parameters.
    }
   {
   We find that there is a close inter-dependency between the thermal evolution and the amount of CO formed, mainly through an important temperature-sensitive CO destruction process with \ch{O+}.
   After a few hundred days, CO completely dominates the cooling of the oxygen-carbon zone of the supernova which, therefore, contributes little optical emission.
   The uncertainty of the calculated CO mass scales approximately linearly with the typical uncertainty factor for individual rates. 
   We demonstrate how molecular masses can potentially be used to constrain various physical parameters of the supernova.
   }
   {}


   \maketitle
%
%________________________________________________________________
%
\section{Introduction}

Molecules in supernovae (SNe) must form in hostile conditions.
The initial supernova explosion releases a large amount of energy ($\sim 10^{51}$ erg for a typical type II-P SN), most of which is converted into kinetic energy that ejects stellar gas at velocities up to a few percent of the speed of light.  
The decay of radioactive species produces a population of high-energy Compton electrons that ionize and dissociate molecules. A strong radiation field in the hot ejecta provides another destruction mechanism.

Despite these destruction mechanisms, there have been molecules detected as early as $\sim 100$ days after the initial explosion. The first molecule identified in any supernova is carbon monoxide (CO) in the nearby SN 1987A at 136 days after the explosion \citep{catchpole_spectroscopic_1988,spyromilio_carbon_1988}, followed by silicon monoxide (SiO) from day 192 \citep{aitken_10_1988,meikle_spectroscopy_1989,roche_silicon_1991}. 
The possible detection of molecules other than CO and SiO at these epochs were also reported, such as \ch{CO+} \citep{meikle_spectroscopy_1989,spyromilio_carbon_1988} and \ch{CS} \citep{meikle_spectroscopy_1989}. 
However, these detections typically relied on a single spectral feature and, therefore, they are more uncertain.
Contemporary observations of SN 1987A have further identified \ch{H2} \citep{fransson_discovery_2016}, \ch{HCO+} , and \ch{SO} \citep{matsuura_alma_2017}, in addition to a continued growth of the CO mass over decades \citep{matsuura_alma_2017}.

Subsequent observations have confirmed that molecular formation similar to that observed in SN 1987A is likely to be commonplace in SNe. 
Both CO and SiO have been identified in several more core-collapse SNe of Type IIP/IIL  \citep{spyromilio_carbon_1996,kotak_early-time_2005,pozzo_optical_2006,kotak_spitzer_2006,kotak_dust_2009,Yuan2016,Banerjee2018,Tinyanont2019}, Type IIb/Ib \citep{Ergon2015,Drout2016}, and Type Ic 
\citep{gerardy_carbon_2002,hunter_extensive_2009}. 
Despite the harsh environment, there seems to be substantial evidence to support a relatively rich molecular chemistry in SNe.
The first models for the creation and destruction of CO and SiO predicted a condensation fraction of the atomic species into molecules on the order of order $10^{-3}$ \citep{petuchowski_co_1989,lepp_molecules_1990}, which is in rough agreement with observations.

The study of molecules in supernovae is important
for several reasons. From molecular rovibrational emission in the infrared, we can infer physical
conditions and supernova parameters, such as temperature and density  \citep[e.g.,][]{liu_oxygen_1995}.
Recently, the Atacama Large Millimeter/submillimeter Array (ALMA) has allowed for this kind of analysis to extend into
the radio regime and for the nearby SN 1987A spatially resolved
maps of the molecules to allow new kinds of constraints to be set \citep{Abellan2017,Cigan2019}.
Forthcoming observations with the James Webb Space Telescope (JWST)  of SNe in the infrared (IR), where the molecular contribution is significant, will also provide a crucial new observational viewpoint.

Beyond their observational signatures, molecules are important for the physical conditions in SNe, which, in turn, govern the atomic emission.
The emission from CO in the IR may dominate the cooling of the oxygen and carbon-rich gas after around a hundred days, which then  affects the ionization state \citep[see e.g.,][]{liu_oxygen_1995}.
Thus, including the effects of molecules on the temperature and ionization is paramount for understanding and modeling the thermal evolution of SNe, which is, in turn, vital for modeling the optical and near infrared (NIR) emission by atoms and ions.

The status of supernova
optical/NIR spectral modeling at nebular phases is now at a level where attempts to
determine detailed nucleosynthesis yields and supernova progenitor masses
are being made \citep[e.g.,][]{dessart_type_2013,jerkstrand_nebular_2014}. 
This has led to a renewed impetus in terms of the red supergiant problem \citep{Smartt2009,Smartt2015} and in gaining an understanding of the landscape of successful and failed explosions as the stellar cores collapse \citep{Oconnor2011,Ertl2016}.
In the majority of such efforts, either the molecules are ignored or treated in a parameterized fashion \citep[e.g.,][]{jerkstrand_progenitor_2012,jerkstrand_nebular_2014}. 

Research that has focused on the molecular formation chemistry \citep[e.g.,][]{petuchowski_co_1989,lepp_molecules_1990,liu_carbon_1992,liu_oxygen_1995,liu_formation_1996,gearhart_carbon_1999} has not considered, on the other hand,   the impact of the molecules on the atomic and ionic emission or the spectral synthesis.
Several works also concentrate on molecule formation as a pathway to dust condensation \citep[e.g.,][]{clayton_condensation_1999,clayton_condensation_2001,cherchneff_chemistry_2009,cherchneff_chemistry_2010,biscaro_molecules_2014,biscaro_molecules_2016,sluder_molecular_2018} given the fact that molecules can be predecessors to dust.
The dust production that is yielded from different types of SNe at different metallicities is still under debate, partly because it is unknown to what extent dust survives shocks produced in later circumstellar interaction stages.
It is unclear how the SN dust yields compare to other major dust production sites, such as the winds of asymptotic giant branch stars and dust growth in the ISM, especially at early times \citep[see e.g.,][]{zhukovska_evolution_2008,valiante_stellar_2009,dwek_origin_2011,triani_origin_2020}.
The main goal of the aforementioned works investigating theories of dust production in SNe is typically to investigate dust formation and provide dust yields for different types of SNe, without necessarily addressing the influence of either molecules or dust on spectra.

The work we initiate here aims to join these two strands (so far) of parallel research and to produce, in a self-consistent way, predictions for
the near-UV to far-infrared (UVOIR) spectra of supernovae that take molecules into consideration.
In this first paper, we describe the methodology used for implementing molecular formation and cooling in the SN spectral synthesis code \textsc{Sumo}, we produce test calculations, which provide estimates of molecular masses, and perform sensitivity tests.
While several different molecular species are present in this first test model, we focus our analysis specifically on CO, as it is known from observations to be abundant in SNe and acts an important coolant, possibly cooling the carbon and oxygen-rich areas of the ejecta by several thousand degrees. 
Before embarking on realistic multi-zone models in later studies, here we study the most important reactions,  any sensitivity to uncertain chemical reaction rates, and the kind of ejecta properties that could be most promisingly diagnosed by observed molecular emission.

This paper is organized as follows.
Section~\ref{sect:meth} presents a description of the spectral modeling methods and the implementation of molecular formation. 
A first test model of the CO production in SN 1987A is presented in Sect.~\ref{sect:coform}. 
The sensitivity of this model to input parameters is investigated in Sect.~\ref{sect:sens} and the possibility of using inferred molecular masses to constrain the density and deposition energy is explored in Sect.~\ref{sect:constr}.
In Sect.~\ref{sect:sumcon}, we provide our summary and conclusions.

\section{Modeling methods}
\label{sect:meth}

\subsection{Supernova model}
The \textsc{ Sumo} code \citep{jerkstrand_44ti-powered_2011,jerkstrand_progenitor_2012} takes a hydrodynamic model for supernova ejecta (density and composition as a function of velocity) as input and calculates the physical conditions and emergent spectra by solving the temperature equation, the statistical equilibrium equations for both ion abundances and level populations, and the radiation field. It is tailored for the post-peak phases and gives snapshot steady-state solutions at any specified epoch. It currently includes 22 elements between hydrogen and nickel, with about 8,000 levels, and 300,000 lines.

\subsection{Chemical model}

\begin{table*}
\caption{Reaction types included which pertain to molecules.
}
\centering
\begin{tabular}{rll}
\hline \hline
\multicolumn{3}{c}{Thermal collision reactions} \\
\hline
\ch{A + B} &\ch{-> AB + h$\nu$} & Radiative association             \\
\ch{A^+ + B }&\ch{-> AB^+ + h$\nu$} & Radiative association of ions             \\
\ch{AB^+ + C} &\ch{ -> A + BC^+} & Ion-neutral exchange\\
\ch{AB^+ + C} &\ch{ -> AB + C^+}& Charge exchange\\
\ch{AB + C} &\ch{ -> A + BC}& Neutral-neutral exchange\\
\hline
\multicolumn{3}{c}{Recombination reactions} \\
\hline
\ch{AB^+ + e^- } &\ch{-> AB + h$\nu$} & Radiative recombination            \\
\ch{AB^+ + e^- } &\ch{-> A + B} & Dissociative recombination         \\
\hline
\multicolumn{3}{c}{Non-thermal reactions} \\
\hline
\ch{ AB + e^-_{C} } &\ch{-> AB^+ + e^- + e^-_{C}} & Ionization by Compton electrons   \\
\ch{ AB + e^-_{C} } &\ch{-> A + B + e^-_{C}} & Dissociation by Compton electrons \\
\hline
\label{tab:rec}
\end{tabular}
\end{table*}

For modeling molecule formation, we incorporate a chemical kinetic description into the \textsc{Sumo} code. 
For the chemical network we use $N_s$ species and $N_r$ reactions, with species referring to all included neutral atoms, atomic ions, neutral molecules, molecular ions, and free thermal electrons.
As chemical reactions occur, number fractions  either decrease (if reactions destroy a species) or increase (if reactions create a species). 
The rate of change of the number density [X] (cm$^{-3}$) of a species X is solved for in steady state as
\begin{equation}
\label{eq:roc}
    \frac{d(\text{[X]})}{dt} = \sum \mathcal{C}_i - \sum \mathcal{D}_i = 0
,\end{equation}
where $\mathcal{C}_i$ and $\mathcal{D}_i$ are the rates of processes that create and destroy the species, respectively. 

The set of equations \eqref{eq:roc}, one for each included species, form a system of coupled, non-linear, algebraic equations, which is solved with a Newton-Raphson scheme. At each epoch, this is done several times in an outer loop where temperature, radiation field, and ionization balance are iterated over. 
The assumption of a steady state is expected to be a good approximation for the range of epochs investigated in this work (150-600d).
This is discussed in \citet{cherchneff_chemistry_2009} and examined in detail in Sect. 4 in \citet{gearhart_carbon_1999}, where relevant reaction timescales are found to be shorter than the dynamic and radioactive timescales.
The validation of the steady-state approximation is also
revealed from the results of our calculations in Sect. \ref{Sect:res}
It should, however, be noted that steady-state does not necessarily hold at later epochs, as noted in \citet{cherchneff_chemistry_2009}. 

Reactions in a chemical network can be categorized according to how many reactants are involved; unimolecular for one reactant, bimolecular for two reactants, and termolecular for three reactants, and so on. 
In the following examples, the quantities Q$_1$ and Q$_2$ refer to the reacting species, and P$_1$ and P$_2$ to the products of a reaction in the chemical network, which could be either molecules, atoms or ions depending on the specific reaction. 
For a unimolecular reaction (u),
\begin{equation}
\label{eq:uni1}
  \ch{ Q_1 -> P_1 + P_2}
,\end{equation}
where the reactant \ch{Q1} turns into the products \ch{P1} and \ch{P2} , for example, as a result of thermal decomposition. The rates of change of each involved species due to this reaction are:
\begin{equation}
\label{eq:uni2}
    R_u = -\left( \frac{d([\text{Q}_1])}{dt} \right)_u =\left(\frac{d(\text{[P}_1])}{dt} \right)_u= \left(\frac{d(\text{[P}_2])}{dt} \right)_u= k_u  \text{[Q}_1]
,\end{equation}
with $k$ being the temperature-dependent rate coefficient, which is specific for each reaction and denoted by a suitable subscript.

Similarly for a bimolecular reaction (b), for example, a collision between two species, \ch{Q1} and \ch{Q2} , creating products, \ch{P1} and \ch{P2}, as
\begin{equation}
\label{eq:bi1}
  \ch{ Q_1 + Q_2 -> P_1 + P_2 }
,\end{equation}
the rates of change are 
\begin{multline}
\label{eq:bi2}
    R_b = -\left( \frac{d([\text{Q}_1])}{dt} \right)_b =-\left( \frac{d([\text{Q}_2])}{dt} \right)_b =\\=\left(\frac{d(\text{[P}_1])}{dt} \right)_b = \left(\frac{d(\text{[P}_2])}{dt} \right)_b = k_b  \text{[Q}_1] \text{[Q}_2].
\end{multline}
Reactions with more reactants, as with termolecular reactions such as a three-body association, are disregarded here as such reactions only become significant at epochs with higher densities than those examined in this work.

The units of the rate coefficients depend on the number of reactants: s$^{-1}$ and  cm$^{3}$s$^{-1}$ for unimolecular and bimolecular reactions, respectively.
The reaction rates at a temperature, $T$ , are typically expressed in an Arrhenius-type form \citep{mcelroy_umist_2013} as:
\begin{equation}
    k(T)= \alpha \times \left( \frac{T}{300~\mbox{K}} \right)^{\beta} \times \exp({-\gamma / T})
,\end{equation}
where $\alpha$, $\beta,$ and $\gamma$ are parameters that are specific to each reaction.

The various reaction types included and the form of the different rate equations are discussed below and summarized in Table \ref{tab:rec}. 
The primary source of rate coefficients used here is the the UMIST Database for Astrochemistry (\citealt{mcelroy_umist_2013}, \href{http://udfa.ajmarkwick.net/}{www.astrochemistry.net}). 
This is supplemented by reaction rate coefficients from \citet{sluder_molecular_2018} and \citet{cherchneff_chemistry_2009}, where reaction networks in supernova remnants were investigated in a way that is similar to the present work.
See Appendix \ref{ap1} for a complete reference list.

\subsection{Included reaction types}

In the following section, we describe the reaction types used in this work. 
Throughout this section, A, B, C, and D are used to represent (neutral) atoms, while AB and BC represent molecules made up by either A and B or B and C.
While the examples here only include diatomic molecules, the notation scheme can be straightforwardly generalized to triatomic or larger molecules.

\subsubsection{Thermal collision reactions}
Collisions between molecules and atoms, or between atoms and atoms, can change both the composition of the gas by forming or destroying molecules and change the ionization state by charge transfer. 

Molecules may form directly by radiative association, where two species, A and B, combine to an initially energized complex.
As the gas densities generally are too low for collisional stabilization, where a collision with a third species carries away energy before the energized complex can dissociate, the main channel is stabilization by radiation \citep{smith_laboratory_2011}.
Then A and B form the new species, AB, and emit a photon as
\begin{equation}
    \ch{A + B -> AB + h$\nu$}
\end{equation}
for two neutral reactants A and B, or
\begin{equation}
    \ch{A^+ + B -> AB^+ + h$\nu$}
,\end{equation}
in the case of an ion and a neutral.

Species can also rearrange through ion-neutral or neutral-neutral interactions.
During ion-neutral interactions, the neutral species is polarized by the electric field of the ion, which induces an electric dipole moment and causes an attractive force between the ion and the neutral.
The outcome of such interaction depends on what is energetically favorable for the involved species. 
One such possible reaction is an ion-neutral exchange of an atomic or molecular ion colliding with a neutral species exchanging one of the components in the molecule as
\begin{equation}
\ch{AB^+ + C -> A + BC^+}
.\end{equation}
A second possibility is ion-neutral charge exchange, which involves ions colliding with a neutral species transferring charge as
\begin{equation}
    \ch{A^+ + B -> A + B^+}
.\end{equation}
Similarly, there can be charge exchange between a molecule and an atom as 
\begin{equation}
    \ch{AB^+ + C -> AB + C^+}
\end{equation}
or between two molecules.
If the neutral species does not possess a dipole moment, the reaction rates of ion-neutral interactions can be assumed to be temperature independent, however, if the neutral has a dipole moment, the average attraction increases for lower temperatures \citep{larsson_ion_2012}. 
Consequently, these reactions are very important for low-temperature environments, such as ISM and late-stage supernovae.

Similarly, in a neutral-neutral exchange, two neutral species collide and react as
\begin{equation}
    \ch{AB + C -> A + BC}
.\end{equation}
Such interactions are typically weakly attractive at larger distances due to van der Waal forces and repulsive at small distances. 
As a result, there is a higher energy barrier to the reactions, with reaction rates increasing with temperature \citep{smith_laboratory_2011}. 
The temperatures in supernovae are, however, sufficiently high and these types of reactions are included in the network.

 Due to low density and comparably low temperatures, we disregard three-body reactions and thermal fragmentation reactions. 
 The CO thermal fragmentation rate \citep[using rate coefficients from][]{appleton_1970} at the temperature and densities here is always at least five orders of magnitude smaller than other important destruction processes.
We allow for up to two reactants and three products in the reactions. 
All collision reactions are then treated as bimolecular reactions, described in Eqs. \eqref{eq:bi1} and \eqref{eq:bi2}, and the rates used for the included collisions can be seen in Table \ref{table:1}.

\subsubsection{Recombination}

Recombination with electrons can recombine an atomic ion as well as recombine or dissociate a molecular ion (or both).
Radiative recombination between an atomic ion and a thermal electron results in neutralization as
\begin{equation}
\label{eq:atomrecomb}
    \ch{A^+ + e^- -> A + h$\nu$}
.\end{equation}
For the case of molecules more possible outcomes of recombination are feasible, such as
\begin{align}
\label{eq:molrecomb}
\ch{ AB^+ + e^- &-> AB + h$\nu$} .\\
\label{eq:molrecomb2}
\ch{ AB^+ + e^- &-> A + B }.\\
\label{eq:dissociativerec}
\ch{ AB^+ + e^- &-> A^+ + B + e^- } .\\
\label{eq:dissociativerec2}
\ch{ AB^+ + e^- &-> A + B^+ + e^- }.
\end{align}
Here, the reaction in Eq.~\eqref{eq:molrecomb} is analogous to the atomic case in Eq. \eqref{eq:atomrecomb}.
For molecules, the dissociative recombination pathway in Eq.~\eqref{eq:molrecomb2} will typically dominate by a large factor \citep{larsson_ion_2012}.
The rates for the molecular dissociative recombination reactions (Eq.~\ref{eq:molrecomb2}) used in this work are listed in Table \ref{table:2}. 
We also assume that radiative recombination reactions of the type shown in Eq.~\eqref{eq:molrecomb} occurs with a small fraction (0.1\%) of the corresponding dissociative recombination reaction rate and that dissociative recombination reactions resulting in ions (Eqs. \ref{eq:dissociativerec}, \ref{eq:dissociativerec2}) are negligible. 

\subsubsection{Destruction by Compton electrons}
Supernovae provide one quite particular component to the physical environment; a population of high-energy Compton electrons (which we label $\rm{e}_{\rm{C}}^{-}$ here) that give a significant and often dominant molecular destruction rate.
These Compton electrons originate from $\gamma$-photons produced in radioactive decay that repeatedly Compton scatter on bound and free thermal electrons, giving them high energies. These high-energy electrons can ionize atomic species, and ionize or dissociate molecules. 
For a diatomic molecule the following destruction processes are considered:
\begin{align}
\ch{ AB + e^-_{C} &-> AB^+ + e^- + e^-_{C}}. \\
\ch{ AB + e^-_{C} &-> A + B + e^-_{C}}. \\
\label{eq:compton1}
\ch{ AB + e^-_{C} &-> A^+ + B + e^- + e^-_{C}}. \\
\label{eq:compton2}
\ch{ AB + e^-_{C} &-> A + B^+ + e^- + e^-_{C}}.
\end{align}

The rate coefficients for these reactions can be calculated by assuming energy from Compton electron collisions is deposited in the zone at a rate of $L$ [erg s$^{-1}$].
For $N_{tot}$ total number of molecules or atoms, the energy deposition rate per particle is then $L / N_{tot}$. 
The rate coefficient for a specific Compton destruction reaction can then be defined as
\begin{equation}
    k_{C} = \frac{L}{N_{tot} W_i}
,\end{equation}
where $W_i$ is the mean energy per ion pair for a given species. It generally depends on the composition and ionization state.
The rate of change per volume for a reaction involving species AB, for example \ch{ AB + e^-_{C} -> AB^+ + e^- + e^-_{C}}, then $R_C = k_C$[AB], effectively treating it as a unimolecular reaction, as described in Eqs. \eqref{eq:uni1} and \eqref{eq:uni2}.

The Compton electron destruction rates for CO were investigated in \citet{liu_oxygen_1995}, presenting a relationship between the electron fraction and $W_{i}$, for a fixed composition (49.5\% O I, 49.5\% C I, 1\% CO I). 
$W_{i}$  settles to a constant value at low values of electron fraction, $x_e \lesssim 0.01$. A constant $W_{i}$ is often assumed \citep[e.g.,][]{cherchneff_chemistry_2009,sluder_molecular_2018}, but 
  instead, we fit a cubic polynomial to the data presented in Fig. 2 in \citet{liu_oxygen_1995}, therefore allowing $W_i$ to vary with the fractional ionization. 
For details about the cubic polynomial fits, see Appendix \ref{polyfit_1}.

We note that SUMO contains a solver for the Compton electron distribution and does calculate the non-thermal destruction rates for atoms and ions. However, adding molecules to this module is a significant task that lies outside the scope of this paper. 
Here, we aim to describe the chemical reaction network and CO NLTE cooling in a test model, so we do not rely on a more detailed treatment of this aspect.

In the case of CO, which is the primary subject of this work's investigation, the branching fractions leading to dissociation into an atomic ion and electron (the reactions in Eqs. \ref{eq:compton1} and \ref{eq:compton2}) are much smaller than the first two reaction types, and are therefore disregarded. 
For other molecules with no available data for $W_i$, we assume the same value as for CO.

\subsection{Reverse reactions}

Between specific internal states, $i$ and $j$, any reaction has a reverse reaction whose rate obeys the detailed balance: $R_{\rm reverse} = R_{\rm forward} g_i/g_f  \exp{\left(-\Delta E/kT\right)}$, where $g$ are the statistical weights and $\Delta E$ the energy difference. The exponential term is for the assumption of a thermal collision partner. In practice, it is often only the total forward rate, to a sum of target states, that is known. This prevents a direct calculation of state-specific reverse rates. Because of this, in this work, we do not attempt to estimate unknown reverse rates for any forward rate. When both forward and reverse rates are presented in the literature, they typically do not obey a detailed balance relation for the same reason (reactants are in the ground state whereas products are in multiple excited states).

While it may not serve as a very accurate description (per the discussion above), we assume, for practical purposes within the network, that products are formed in their ground states. For molecules that are not explicitly treated as multi-level NLTE elements, this is, in fact, the only state. We further assume, as per the standard in \textsc{SUMO}, that only elements in their ground multiplet state contribute to a chemical reaction (but this fraction is typically close to unity).

\subsection{Photoionization}
No work that we are aware of to date  has calculated self-consistent photoionization rates for molecules in supernova ejecta. While this is something we hope to do in subsequent work, we do not attempt it here (whereas photoionization for atoms and ions is considered as is customary in \textsc{Sumo}). 
We note that thanks to the use of approximative rates, it is estimated that the UV destruction of molecules is a sub-dominant process \citep{cherchneff_chemistry_2009}.

\subsection{CO line emission cooling}

The formation of CO is expected to affect the temperature evolution of the SN ejecta, as CO emission in the infrared adds a new efficient cooling channel. 
A previous work investigating CO cooling in SNe has indicated that the effect is substantial; \cite{liu_oxygen_1995} found that CO emission could lower the temperature by several thousand degrees in the O/C zone. 

The NLTE cooling by CO ro-vibrational transitions was added into SUMO, adopting energy levels and transition probabilities from \citet{li_rovibrational_2015}, including levels up to the vibrational quantum number $\nu = 6$ and rotational quantum number $J = 32$\footnote{This data is available through the online material of \citet{li_rovibrational_2015}}. Tests showed that this was sufficient, whereas the inclusion of more states did not significantly influence the cooling from CO.  Ro-vibrational radiative transitions for the fundamental, first, and second overtones were included. Pure rotational transitions were ignored, as they are not important for the cooling due to the low transition probabilities and small energies involved.
Optical depth for the molecular line emission is considered, as it is for the atomic lines, by the Sobolev formalism.

Collision rates were treated with the standard default approximation in \textsc{Sumo}, which uses (for transitions with no measured or calculated values) a collision strength of $\Upsilon=0.004g_1g_2$ for forbidden transitions (and van Regemorter's formula for allowed). Recombination was ignored in calculating the level populations.

\section{Test case: CO formation in SN 1987A}
\label{sect:coform}

\begin{table}
\caption{Neutral atoms, atomic ions, neutral molecules, and molecular ions included in the simulations. }             % title of Table
\label{tab:spec}      % is used to refer this table in the text
\centering                          % used for centering table
\begin{tabular}{l l}        % centered columns (4 columns)
\hline\hline                 % inserts double horizontal lines
Type & Species \\    % table heading 
\hline                        % inserts single horizontal line
Neutral atoms & C, O \\
Atomic ions & C$^+$, C$^{2+}$, O$^+$, O$^{2+}$ \\
Neutral molecules & C$_2$, O$_2$, CO, CO$_2$, C$_2$O \\
Molecular ions & C$_2^{+}$, O$_2^{+}$, CO$^{+}$, CO$_2^{+}$, C$_2$O$^{+}$ \\ 
\hline                                   %inserts single line
\end{tabular}
\end{table}

To validate the modeling methods and to perform sensitivity tests, we explored CO formation in a simplified one-zone model of O/C-composition for SN 1987A, devoid of any elements other than oxygen and carbon. Since most CO likely forms in the O/C-zone \citep{liu_oxygen_1995}, our calculated CO masses for this test model should be comparable to observationally inferred yields from SN 1987A. 
The model is based on the O/C zone of a $M_{\rm ZAMS}=19$ \msun~core-collapse explosion model of SN 1987A, originally derived from models presented in \citet{woosley_nucleosynthesis_2007} and previously investigated using \textsc{Sumo} in \citet{jerkstrand_44ti-powered_2011,jerkstrand_progenitor_2012}. 

This zone has a mass of 0.58 M$_\odot$ and mass fractions of 0.62 O, 0.28 C, 0.078 He, 0.018 Ne, and 4.9$\times 10^{-3}$ Mg, plus traces of other elements. We note that the He abundance is typically negligible in the C/O zone, thus, the large value here of 8\% is an outlier among the 12, 15, 19, and 25 \msun~models.
As we only include C and O here, we normalize their mass fractions to 0.69 and 0.31, respectively. 
The zone was given a spherical shape, with $V_{\rm in}=0$ \kms~and $V_{\rm out}=540$ \kms. 
This gives it the same density as in \citet{jerkstrand_progenitor_2012}; in that more realistic multi-zone model, the material resides as multiple clumps over a larger velocity span.  
While ignoring the other SN zones would change the radiation field in the zone, the (ionizing) UV radiation is mostly emitted locally. 
In addition, as mentioned previously, in this first test model we treat photoionization only for atoms and ions but not for molecules.
We adopt the gamma deposition calculation from \citet{jerkstrand_progenitor_2012} (see Appendix \ref{ap:depe} here). The species included in this step are listed in Table \ref{tab:spec} and the reaction network used can be seen in Appendix \ref{ap1}.
We model the supernova between 150 and 600 days, with time intervals of 50 days.

\subsection{Results}
\label{Sect:res}

   \begin{figure}
   \centering
   \includegraphics[width=\hsize]{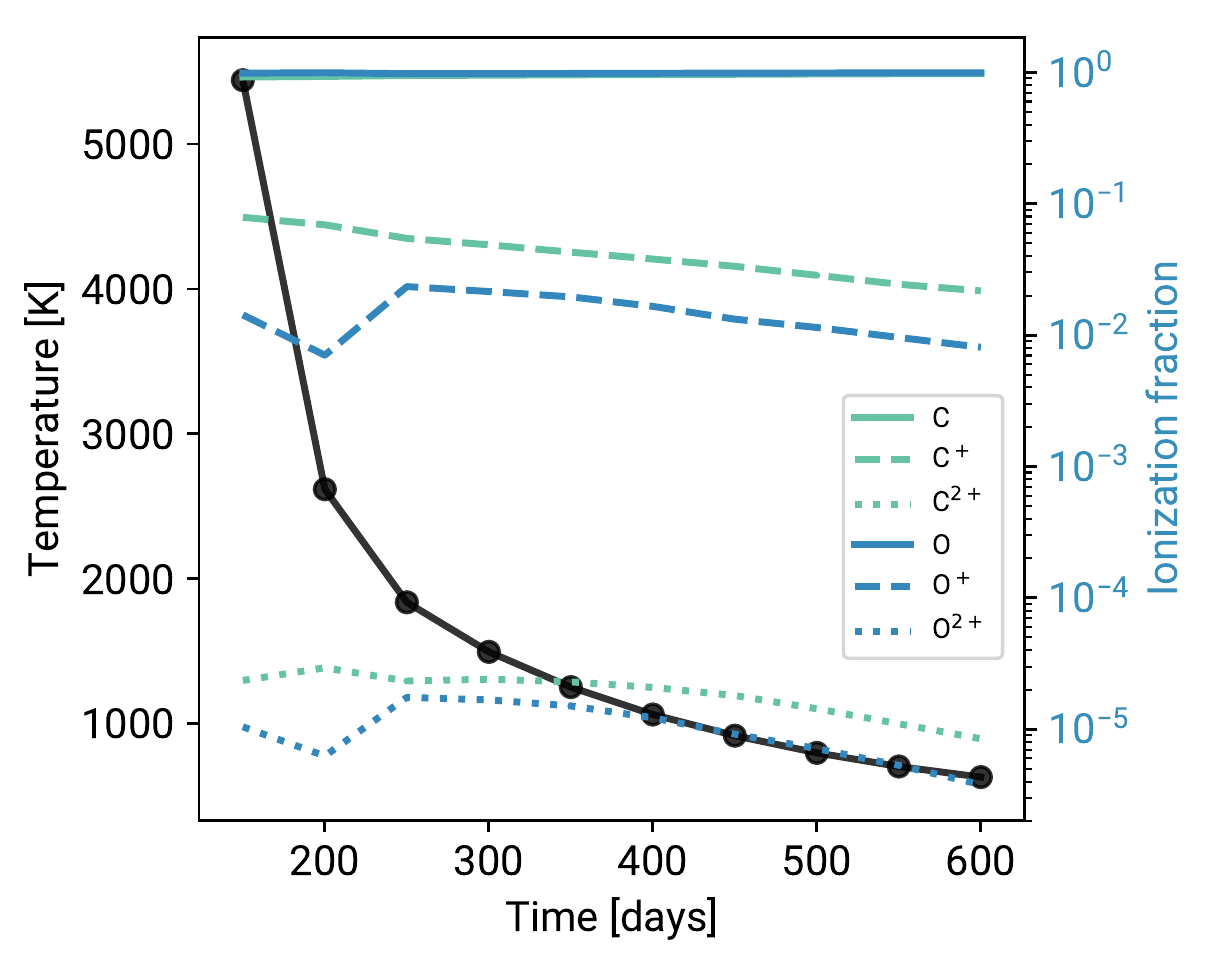}
      \caption{Black line shows the temperature (left axis) of the SN 1987A model and the colored lines show the ionization degrees (right axis) of carbon (teal) and oxygen (blue). 
      }
         \label{fig:ionfracs}
   \end{figure}

   \begin{figure}
   \centering
   \includegraphics[width=\hsize]{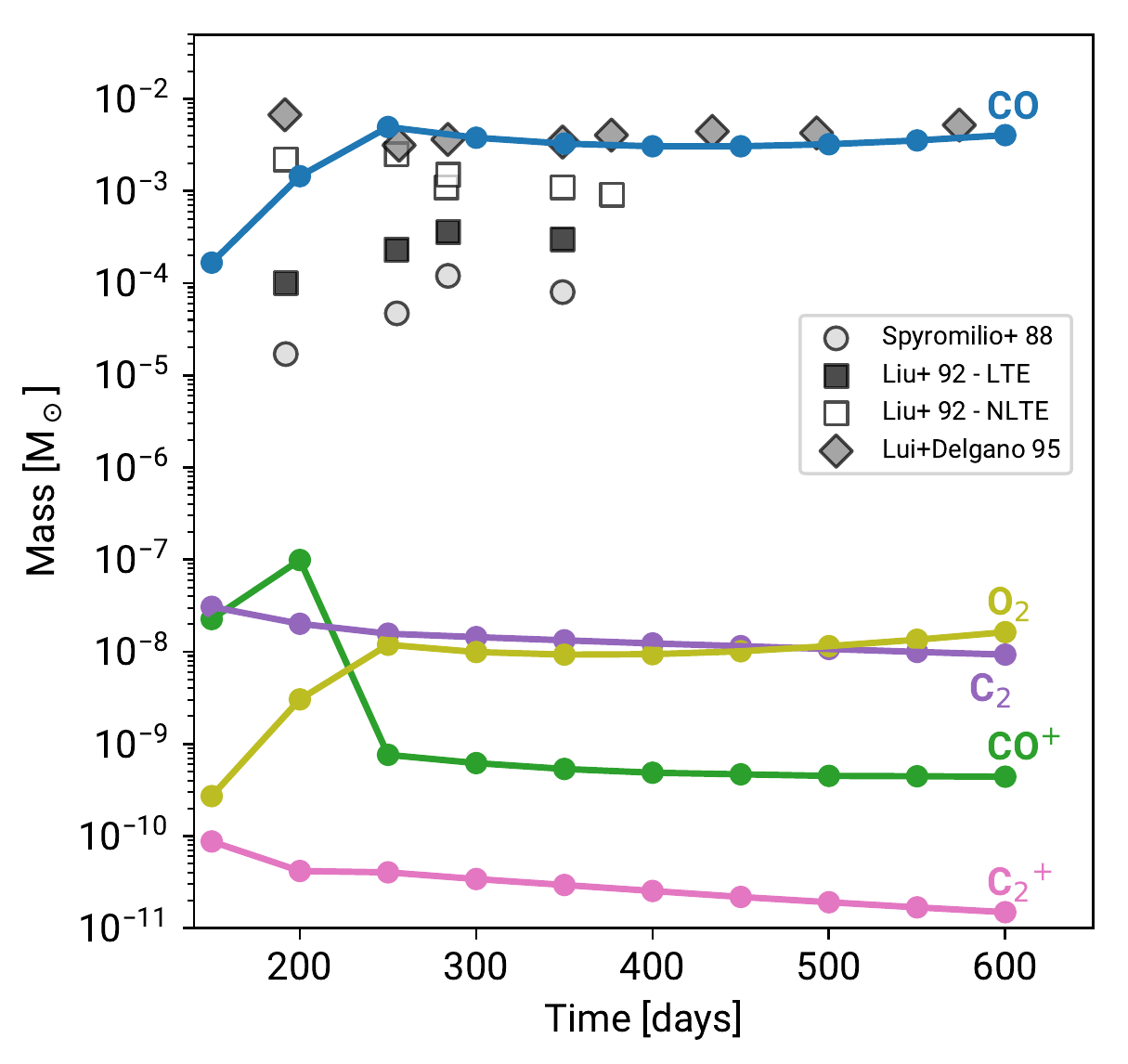}
      \caption{Lines represent molecular masses from our model over time and the scatter points are observational estimates of the CO masses. The observational estimates using NLTE models with optical depth consideration (white squares, \citealt{liu_carbon_1992}; and grey diamonds, \citealt{liu_oxygen_1995}) generally give values that are close to the model prediction.}
         \label{fig:co_res1}
   \end{figure}

% \fixme{make table with days and temperatures, and maybe ionization fractions?}

The resulting temperature and ionization fractions of C and O in the model are shown in Fig. \ref{fig:ionfracs}.
Over the time we investigated, the temperature drops from around 5000 K down to 800 K.
The temperature drops significantly when the amount of CO starts to become significant, which is at around 200d (Fig. \ref{fig:co_res1}).
There is initially around 10\% singly ionized carbon and around 1\% singly ionized O in these models, which is decreasing with time. 
We note that the kinks seen at 200 days, in the time evolution of \ch{O+} and \ch{O^{2+}}, are due to a temperature-dependent reaction involving \ch{O+} (Eq.~\ref{eq:oplus}) that turns off at lower temperatures ($\lesssim 2000$ K), thus changing the trends seen (see also discussion in Sect. \ref{sect:creation_paths}).
The fractions of doubly ionized C and O are orders of magnitude smaller and these play no role in the physical conditions. For CO, the degree of ionization is much smaller than for the atoms, with CO$^+$/CO $\lesssim 10^{-4}$. 

In Fig. \ref{fig:co_res1}, the molecular masses against time are shown, with the most abundant molecule being CO. At the first epoch of 150d only a relatively small amount of $\sim 10^{-4}$ \msun~forms. There is then a rapid increase up to about 250d, where the CO mass levels out at $ \sim 4 \times 10^{-3}$ M$_\odot$.
Other molecular species only form in small amounts compared to CO, with masses several orders of magnitude lower. 
Species not plotted only formed in minuscule amounts.

In Fig. \ref{fig:co_res1} also observational estimates of CO masses for SN 1987A are plotted \citep{spyromilio_carbon_1988,liu_carbon_1992,liu_oxygen_1995}. 
These estimates are derived from the same infrared observations of CO first overtone bands, however, several different assumptions are used and the reported CO mass varies by 1-2 orders of magnitude between the different works, indicating significant model dependency.
In the initial paper by \citet{spyromilio_carbon_1988} LTE and optically thin conditions were assumed, and they derived a CO mass of $\sim 10^{-5}-10^{-4}$ M$_\odot$, with a tendency of growth from 200d to 350d.
Later \citet{liu_carbon_1992} found a larger CO mass ($ \sim 10^{-3}$ M$_\odot$) when considering NLTE and optical depth effects; both effects were found to give a too low CO mass if ignored. In their NLTE model no time-dependence of the CO mass was seen, or a slight decrease with time.
\citet{liu_oxygen_1995} took a somewhat different approach, using simulated model temperatures instead of having temperature as a free fitting parameter as in the previous works.
With this method, they found a slightly higher mass than in \cite{liu_carbon_1992} ($5 \times 10^{-3}$ M$_\odot$), with significantly cooler temperatures than from the free-fitting approach. New data allowed the range of epochs to be extended to 550d; also, they found no time evolution in the inferred CO mass. 
The papers where optical depth is taken into account \citep{liu_carbon_1992,liu_oxygen_1995} generally produce better spectral fits to the data.

Keeping these differences in mind, as they result in a significant spread in the observational estimates of CO masses, there is a general agreement between the observations and our model results.
Between the optically thin LTE results and our models, there are between one and two orders of magnitude of difference, however, for the NLTE models, the difference is less than 0.5 orders of magnitude.
Our results are notably similar to the estimates in \citet{liu_oxygen_1995} after 250 days.

With our standard SN 1987A model, we come to the same conclusion as \citet{gearhart_carbon_1999} and \citet{cherchneff_chemistry_2009} concerning the validity of using a steady-state solution for a small network. 
If we define a characteristic timescale of a process as the ratio of the quantity divided by the rate of change of that quantity, $\tau = \phi / \dot{\phi}$, we can estimate the dynamical timescale as $\tau_{dyn} \approx n / \dot{n} \approx 1/3 \times t$ and the radioactivity timescale as the decay time of \ch{^{56}Co}, i.e. $\tau_{rad} = \tau_{^{56}Co} = 111 \text{ days}$.
At a few hundred days post explosion, meaning that at times typical to nebular phase supernovae, these two timescales are comparable (e.g., at 300 days, $\tau_{dyn} \approx \tau_{rad} \approx 100 \textrm{ days}$). 
In contrast, the timescales of the important chemical reactions $\tau_{chem}$ tend to be significantly shorter.
If we take $\tau_{chem}$ as $ [\text{X}] / \sum{ \mathcal{D}_i}$, where $\sum{ \mathcal{D}_i}$ is the sum over all the the destruction rates of X, we obtain typical values of $\tau_{chem} \approx 10^{-2}\textrm{ days}$ at 200 days for CO, monotonically increasing to $\tau_{chem} \approx 4\textrm{ days}$ at 600 days. 
Because $\tau_{chem} \ll \tau_{dyn},\tau_{rad}$ the steady state approximation is valid.

\subsection{The principal paths for CO}
\label{sect:creation_paths}

\begin{figure*}
\centering
\includegraphics[width=0.49\textwidth]{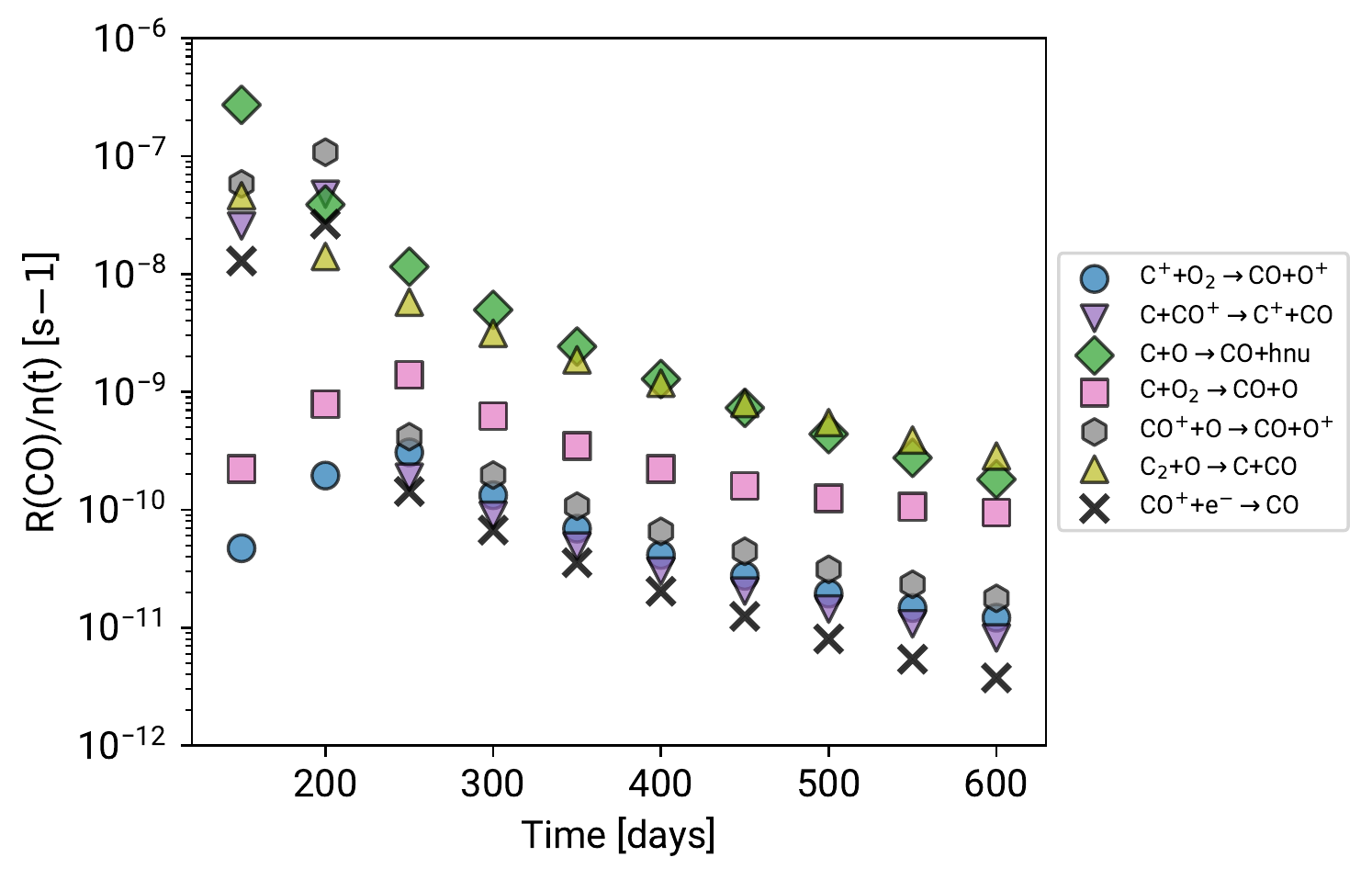} 
\includegraphics[width=0.49\textwidth]{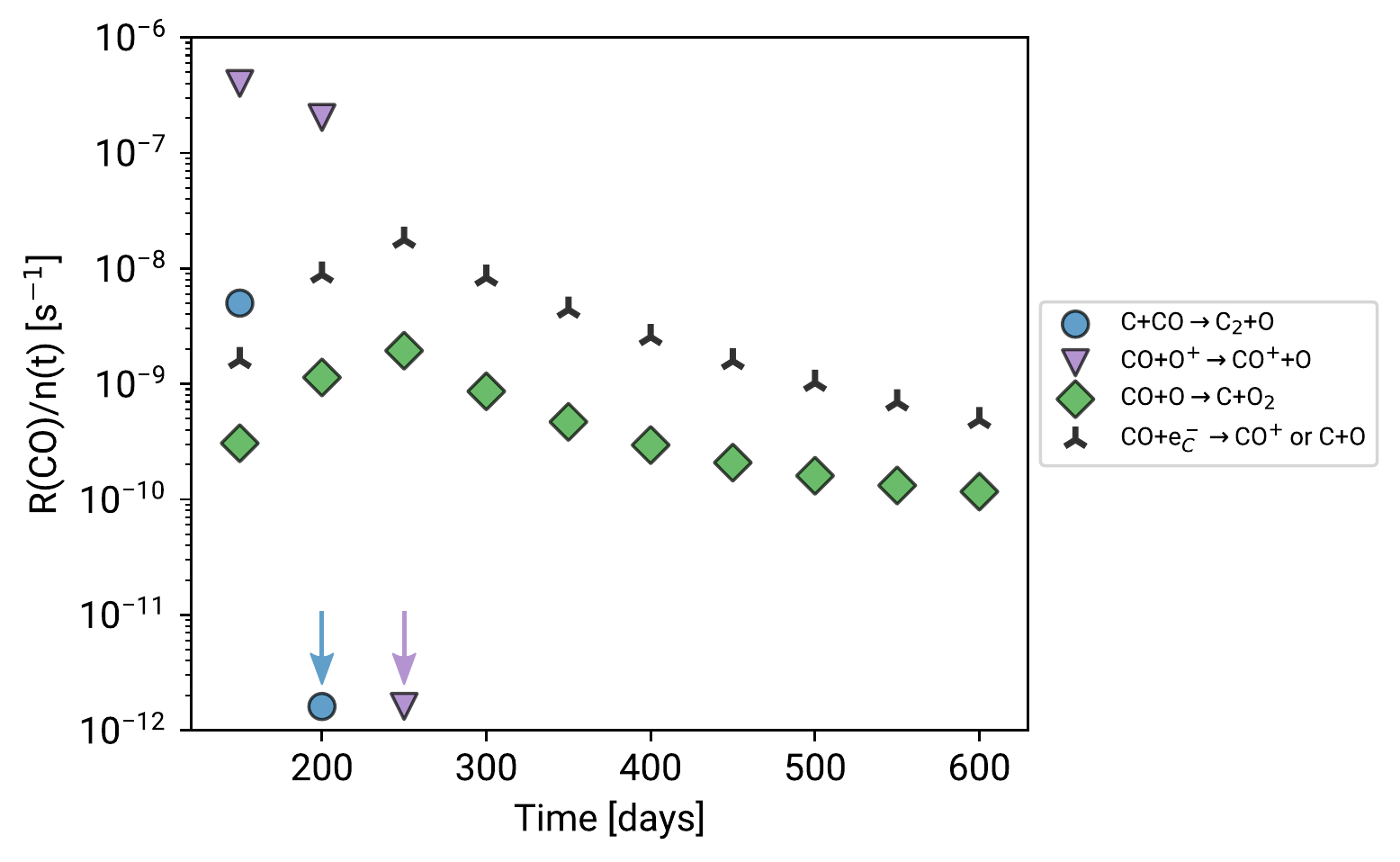}
\caption{Reactions important for the formation and destruction of CO over time for the SN 1987A model. \textit{Left -} Rates of different reactions creating CO divided by the total number density. \textit{Right -} Destruction rates of CO divided by the total number density. }
\label{fig:rate_res}
\end{figure*}

The panels of Fig. \ref{fig:rate_res} show the reaction rates (i.e., either $R_u$ or $R_b$ depending on the reaction, as described in Eqs. \eqref{eq:uni2}, \eqref{eq:bi2}) for the most important processes divided by the total number density at that time-step. 
This gives rates per particle (unit s$^{-1}$) which removes the $t^{-3}$ background evolution of rates per unit volume due to the homologous expansion. We note that destruction rates are not per CO particle but per any particle; to get the result per CO particle, we divide by the number fraction of CO.
The processes creating CO are shown in the left panel and processes that destroy CO are shown in the right panel. 
As seen in the left panel of Fig. \ref{fig:rate_res}, important creation pathways for CO at all times are the radiative association between C and O as
\begin{equation}
    \ch{C + O -> CO + h$\nu$},
    \label{eq:raco}
\end{equation}
and neutral exchange with \ch{C_2}
\begin{equation}
    \ch{C_2 + O -> C + CO.}
\end{equation}
The \ch{C_2} here forms mainly through radiative association as
\begin{equation}
    \ch{C + C -> C_2 + h$\nu$.}
\end{equation}

At early times, up to around 200 days, reactions involving \ch{CO^+} have among the highest flows of all CO formation channels. Important reactions are the charge transfer between \ch{CO^+} and O,
\begin{equation}
    \ch{CO^+ + O -> CO + O^+,}
\end{equation}
the analogous charge transfer between \ch{CO^+} and C
\begin{equation}
    \ch{CO^+ + C -> CO + C^+}
\end{equation}
and the recombination reaction 
\begin{equation}
    \ch{CO^+ + e^- -> CO.}
\end{equation}
At these times, \ch{CO^+} is in turn primarily created by radiative association between \ch{C} and \ch{O^+}, and by charge exchange between \ch{CO} and \ch{O^+}. 
These channels are, however, effectively quenched at later times as the latter reaction 
\begin{equation}
    \ch{CO + O^+ -> CO^+ + O}
    \label{eq:oplus}
\end{equation}
is endothermic with a few tenths of an eV and turns off when the gas gets too cool
(the rate is derived and discussed in \citealt{petuchowski_co_1989} based on cross section measurements from \citealt{murad_reaction_1973}). This has two consequences; as seen in Fig. \ref{fig:co_res1}, the amount of \ch{CO^+} decreases significantly between 200 and 250 days, as one of the major creation channels is turned off. 
The CO creation pathways involving \ch{CO^+} are suppressed as a result, as seen in the left panel of Fig. \ref{fig:rate_res}.
Secondly, the dominating destruction channel for CO at early epochs turns off. 
The combined effect is a rapid increase in CO mass. 
While there are two other CO creation pathways of a similar efficiency as the ones involving \ch{CO^+}, the rate in Eq.~\eqref{eq:oplus} strongly dominates the other destruction processes.
When it turns off, the destruction rate, therefore, is reduced by a larger factor than the creation rate, and the CO mass increases. The evolution of \ch{CO} mass is consequently intimately related to both the formation and destruction of other molecules and to the temperature in a complex way. 
We note that, in general, the full impact of any atom or molecule in the reaction network for the CO abundance (or any other abundance) cannot be fully established from plots like these. 
It would be necessary to run the network with that particle extracted. However, if a rate is large, the reaction is potentially important and should be included.

The two dominant destruction pathways from 250 days onwards are the destruction by Compton electrons as
\begin{equation}
    \ch{CO + e^-_C -> CO^+} \text{ or } \ch{C + O}
\end{equation}
and the neutral exchange with O as
\begin{equation}
    \ch{CO + O -> C + O_2}.
\end{equation}

These results broadly agree with the findings of previous works concerning CO formation in supernovae. 
There is a general consensus that the radiative association between C and O is the most important creation pathway for CO in this type of supernova \citep[see e.g.,][]{lepp_molecules_1990,liu_carbon_1992,gearhart_carbon_1999,sarangi_chemically_2013}. 
The importance of other formation processes discussed here also tend to be mentioned.
It should be noted, however, that these results are dependent on assumptions about the physical conditions as well as the rate coefficients used in the network. 
The self-consistent temperature calculation shown here represents one major improvement in reducing free parameters in the modeling.

\subsection{Thermal evolution}

   \begin{figure}
   \centering
   \includegraphics[width=\hsize]{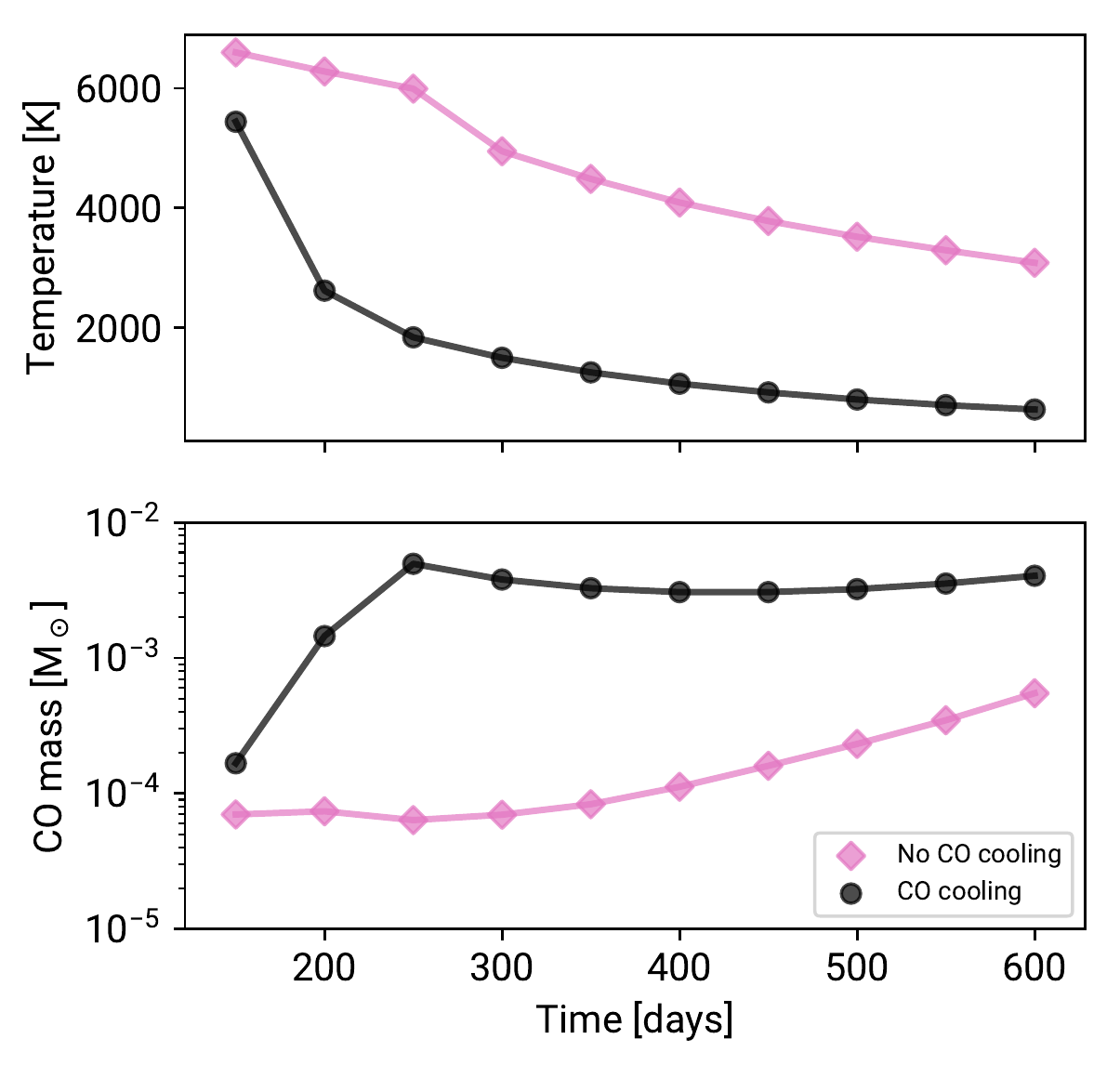}
      \caption{\textit{Upper -} Temperature evolution of the SN 1987A model, with (black) and without (pink) CO cooling. \textit{Lower -} Mass of CO formed over time for the SN 1987A model, with (black) and without (pink) CO cooling. 
      }
         \label{fig:coolvsnocool}
   \end{figure}

There is a complex interplay between the temperature and the CO mass. 
The CO effectively cools the gas which, in turn, affects some of the temperature-sensitive chemical reaction rates that govern the creation and destruction of CO. 
As mentioned in the previous section, the relationship between CO mass and temperature is non-trivial, however, a cooler environment generally leads to more CO being created. 
The main destruction channel at higher temperatures, described in Eq.~\eqref{eq:oplus}, is very temperature-dependent and quickly becomes inefficient when the temperature drops. 
Once CO starts to form,  there is positive feedback; CO then  helps to cool the gas by its rovibrational emission and with a cooler temperature, more CO will form.
This sustains until critical temperature-dependent reactions, such as charge transfer destruction with O$^+,$ turn off.

This can be seen in Fig. \ref{fig:coolvsnocool}, where the temperature of our SN 1987A O/C-zone model has been calculated with and without CO cooling included. 
The effect is substantial; when CO cooling is included the temperature is several thousand degrees lower and the amount of CO formed about an order of magnitude larger.
This indicates that even a small amount of CO has large consequences for the thermal evolution.
These results broadly agree with previous findings by \citet{liu_oxygen_1995}.

\subsection{Comparison to other works}

Several theoretical investigations into the CO production in SNe have been carried out since the first detection in SN 1987A.
Table \ref{table:workcomp} in Appendix \ref{ap:workcomp} shows an overview of the CO mass reported in selected works that specifically model CO masses which are comparable to those in SN 1987A, much as in this paper, at 300 days.

There is a large spread in results; early works typically found a CO mass of 
$\sim 10^{-5}-10^{-6}$ \msun~ \citep[e.g.,][]{petuchowski_co_1989,lepp_molecules_1990}.
Modeling in \citet{liu_oxygen_1995}, where the cooling effects of CO were taken into account, gave a higher CO mass of $\sim 10^{-3}$ \msun, while more recent works \citep[e.g.,][]{sarangi_chemically_2013,sluder_molecular_2018} find a CO mass of $\sim 10^{-2}$ \msun.
Our results ($4 \times 10^{-3}$ \msun~at 300 days) generally agree well with \citet{liu_oxygen_1995} and are within an order of magnitude to the findings in \citet{sarangi_chemically_2013} and \citet{sluder_molecular_2018}.

Differences in networks and reaction rates used and the inclusion of processes such as photoionization, cooling by CO, and Compton destruction in some works are possible explanations for the differences.
There are also other possible reasons for the disparity, such as different modeling parameters, for example, the mass and density of the O/C zone, the fractions of O and C, etc.
Table \ref{table:workcomp} also compares some key modeling parameters and, as we can see, there are significant differences between the mentioned quantities;
there is up to a factor five difference between the C and O masses, an order of magnitude difference in number densities, and up to 2500 K difference in temperatures (mostly parameterized rather than calculated).

Earlier works \citep{petuchowski_co_1989,lepp_molecules_1990,liu_carbon_1992} typically also assumed a fully mixed model, with several solar masses of helium (He), making the CO destruction by collision with \ch{He+} very efficient. 
For in-depth discussions on the differences between earlier theoretical works, see \citet{gearhart_carbon_1999} and \citet{sluder_molecular_2018}.

\section{Sensitivity tests}
\label{sect:sens}

Here, we investigate how sensitive the resulting molecular masses are to modeling assumptions and uncertainties, which roughly can be divided into two categories; assumptions about the physical properties of the supernova (e.g., deposition energy, density, composition,..) and uncertainties of the chemical network, that is, the rate coefficients. 

\subsection{Varying the physical parameters}
\label{sect:phys_par}

\begin{figure*}
\centering
\includegraphics[width=0.49\textwidth]{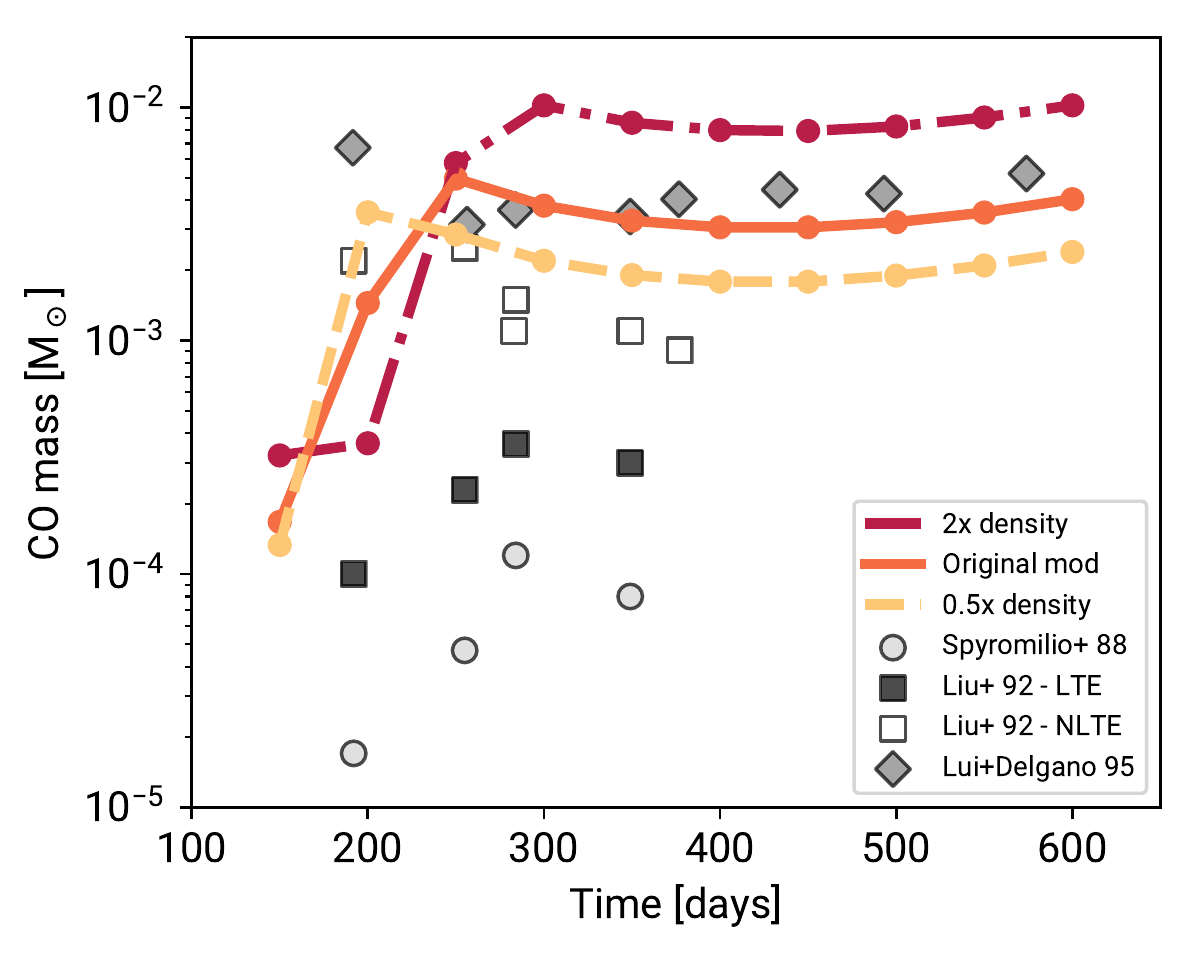} 
\includegraphics[width=0.49\textwidth]{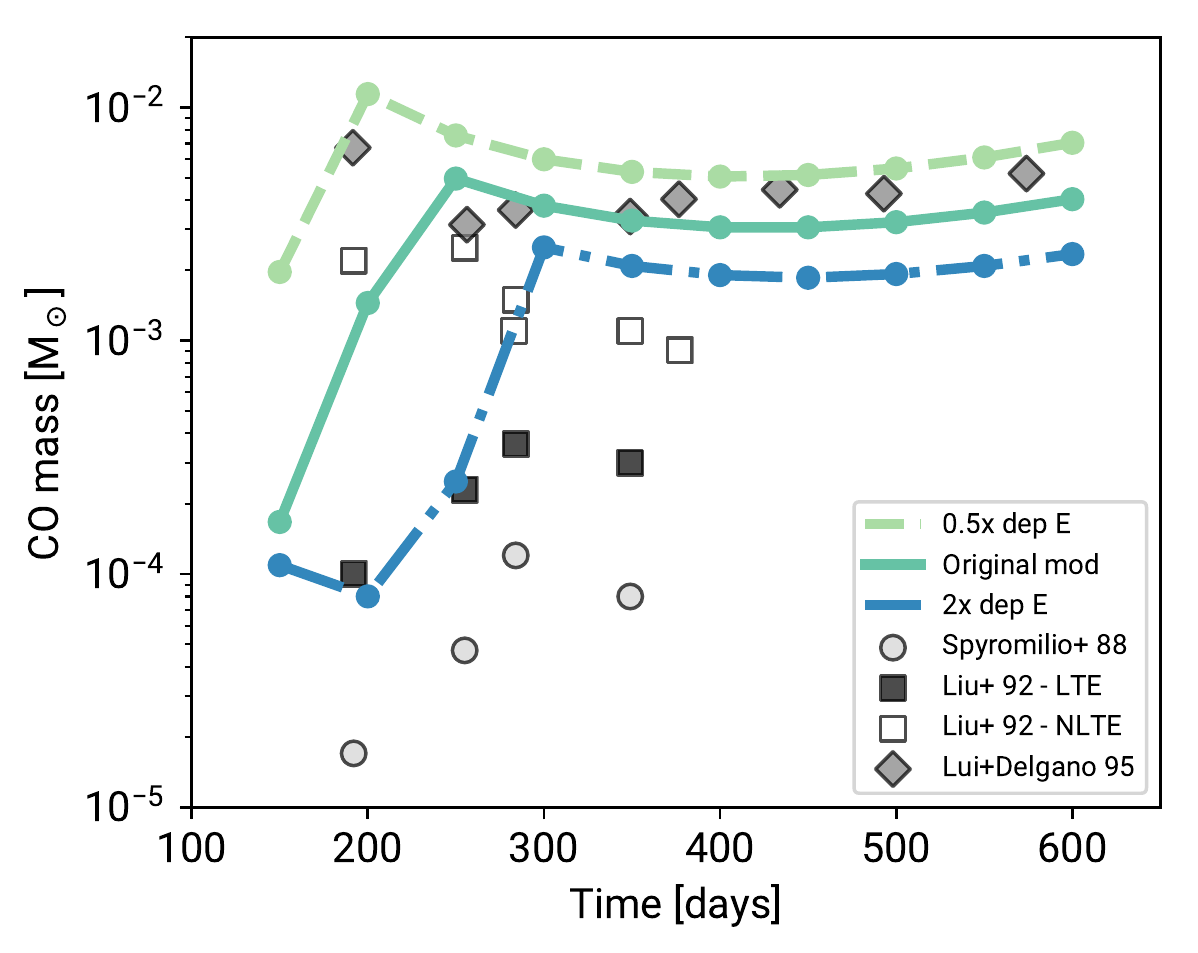}
\caption{Results of varying the physical parameters of the SN1987A model. Scatter points are observational estimates. \textit{Left -} Lines show CO mass of models with original density, twice and half the original density. \textit{Right -} Lines show CO mass of models with original deposition energy, twice and half the original deposition energy. }
\label{fig:co_res2}
\end{figure*}

We tested the sensitivity of CO formation on the input parameters density and deposition energy, with Fig. \ref{fig:co_res2} showing the time evolution of the CO mass for different densities in the left panel, and different deposition energies in the right panel. 

For different densities, the models show a complex behavior between 150 and 250 days. A rapid onset of CO formation, seen for the original model and the low-density model between 150 and 200 days, starts later in the high-density model. 
This is due to the previously discussed delicate relationship between the temperature and the CO destruction path through charge transfer with \ch{O^+} (Eq. \ref{eq:oplus}).
The dense model has a higher temperature and it takes a longer time for this model to cool to temperatures where the aforementioned charge transfer reaction turns off. 

After 250 days, there is a positive correlation between density and CO mass, which is more or less uniform over time.
In a symmetric network (with the same number of reactants of all processes), the increase in number density should not significantly change the mass of the formed molecules. 
This is, however, not the case for CO. 
As discussed in Sect. \ref{sect:creation_paths}, several bimolecular processes contribute significantly to the creation of CO, while destruction is dominated by only one process, namely, destruction by Compton electrons, which is treated as a unimolecular process.
While unimolecular processes roughly depend linearly on the number density, bimolecular processes have a square relationship. 
Therefore, if number density is increased, the sum of the creation processes will increase by a larger factor than the sum of the destructive processes in the case of CO, resulting in a higher CO mass. 
Related to this point is the fact that the number density of Compton electrons will stay roughly constant upon compression.

In contrast to density, increased deposition energy decreases the CO mass, as seen in the right plot of Fig. \ref{fig:co_res2}.
When increasing the deposition energy in the model, the temperature and the ionization fractions increase. 
Initially, as the main destruction channel for CO depends on \ch{O^+} (Eq. \ref{eq:oplus}), a larger abundance of \ch{O^+}, as well as a higher temperature, will increase the rate at which CO depletes. 
The main creation channels for CO, on the other hand, do not depend on any atomic ions, so increases in these will have little impact on the total creation rate. 
At later times the destruction by Compton electrons takes over as the dominant destruction channel. 
The Compton electron population increases with higher deposition energy 
and the outcome of increasing deposition energy is, therefore, a more efficient destruction of CO by Compton electrons.

It should be noted that both the positive correlation between model density and CO mass, and the negative correlation between deposition energy and CO mass, are not general results for all molecules, but rather a consequence of the specific creation and destruction channels for CO.
It is most likely that such relationships are unique for each molecule
and need to be investigated independently for each molecule of interest. 

\subsection{Impact of uncertain rate coefficients}

   \begin{figure*}
   \centering
   \includegraphics[width=0.49\textwidth]{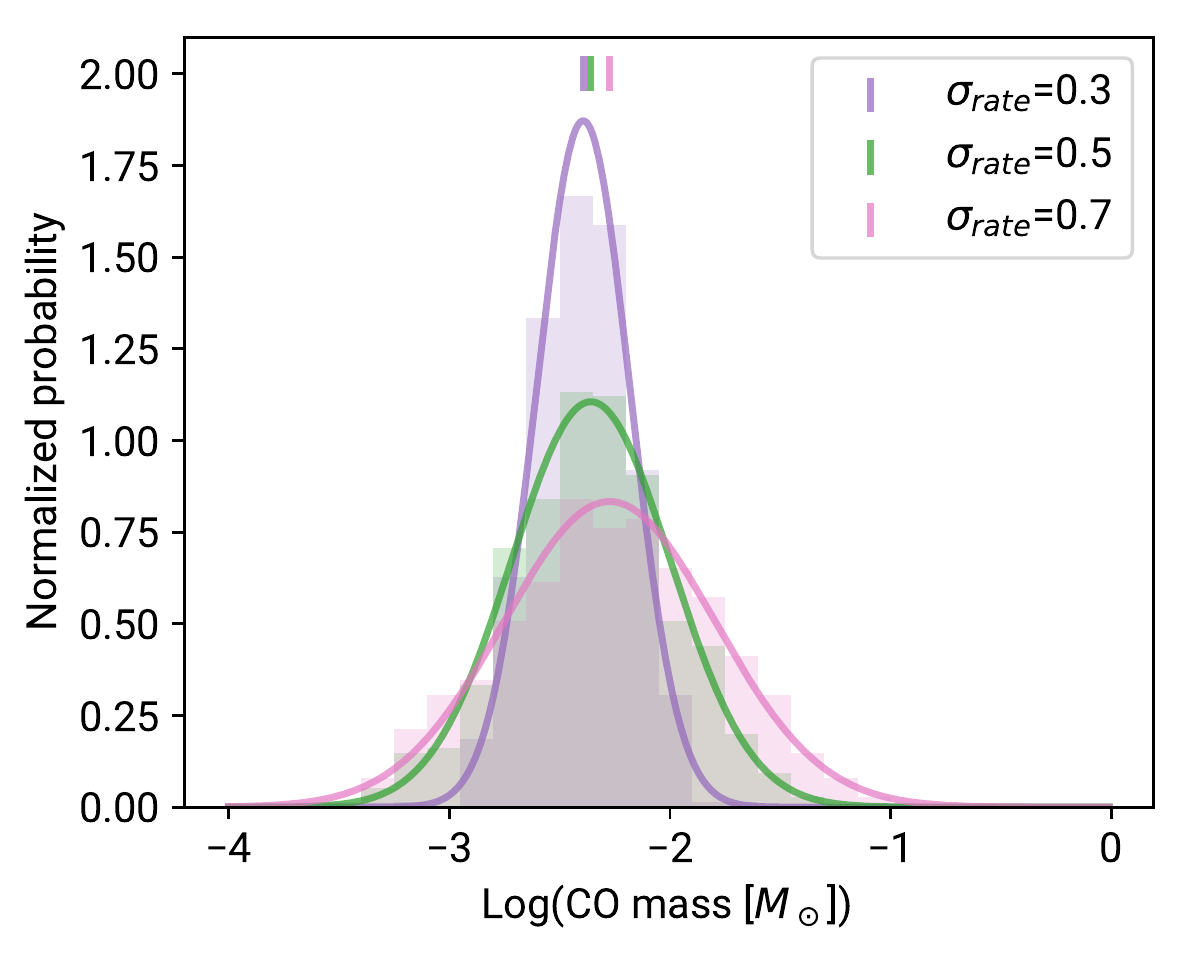}
   \includegraphics[width=0.49\textwidth]{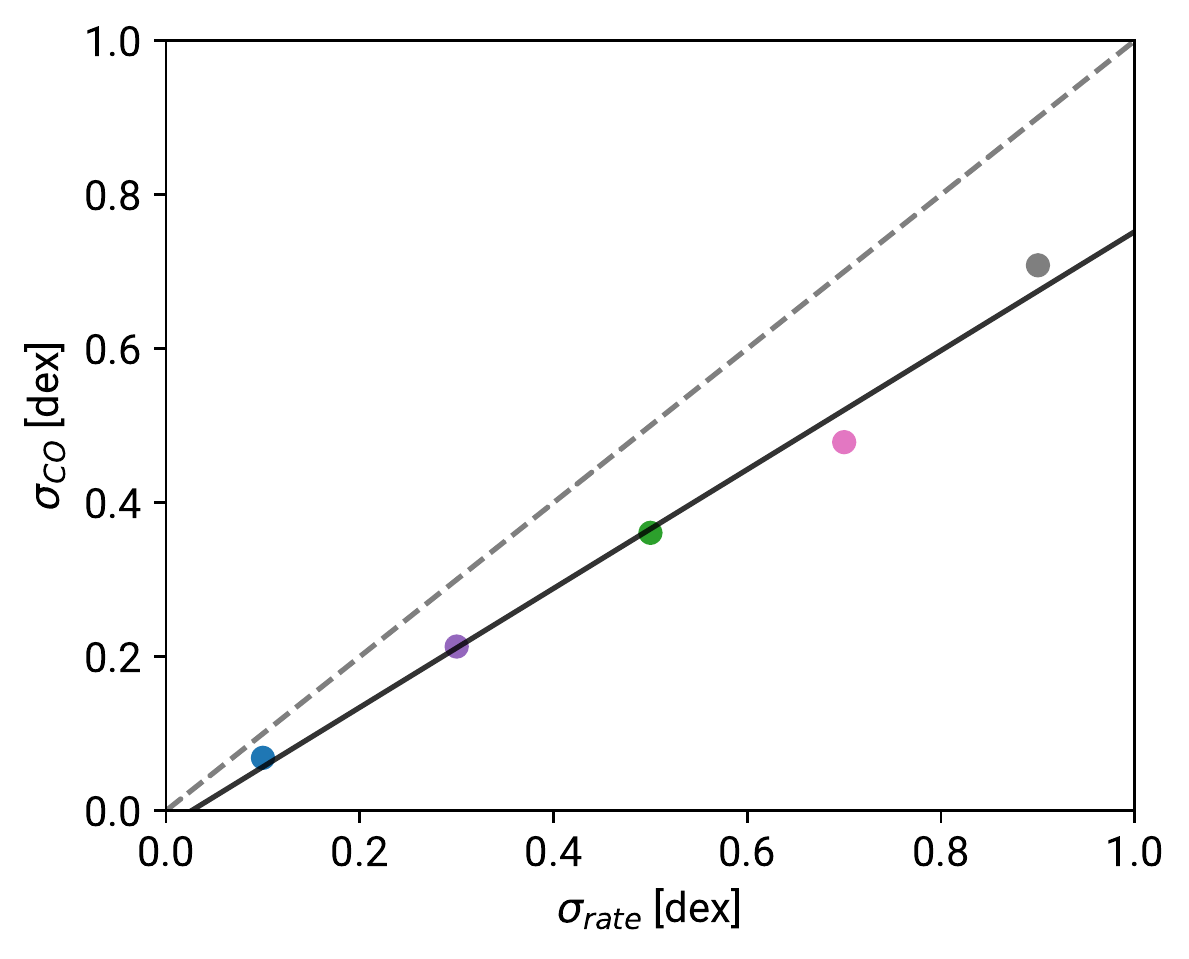}
      \caption{\textit{Left -} CO mass distribution, for the MC tests with three different $\sigma _{rate}$. 
      \textit{Right -} $\sigma_{CO}$ (standard deviation of the resulting CO distribution) against the $\sigma_{rate}$ (standard deviation of the individual rates), for details see text. The solid black line is a fit to the data (slope=0.8), the dashed line is the one-to-one relationship. For more details, see the text. }
         \label{fig:mc1}
   \end{figure*}

The rate coefficients used for describing the thermal collisions involving CO come from multiple sources and are typically derived using different methods (experiments, calculations, or estimates) with a range of uncertainties. 
We want to investigate how changing the thermal collision rate coefficients by some typical uncertainty factor affects the resulting CO masses, for example, to test the degree of robustness with respect to uncertain reaction rates. 
This is done using a Monte Carlo approach.

\subsubsection{Monte Carlo simulations}

% We adopt the following method to assess the impact of changing the thermal collision rates:
% \LEt{ Please consider reworking these points into the text.}\begin{enumerate}
%     \item Each thermal collision rate in the network (as seen in Table~\ref{table:1}) is given a log-normal probability distribution, with a mean equal to the literature value used in the standard network and with a standard deviation $\sigma_{rate}$ (units of dex).
%     \item Each rate coefficient is then randomly sampled from this distribution to make a randomly sampled network.
%     \item The CO mass is solved for using a full \textsc{Sumo} run at 300 days using the standard SN 1987A single-zone model for this randomly sampled network.  
%     \item Steps 1-3 are repeated 500 times, re-sampling the thermal collision rates at each instance to investigate the parameter space.
%     \item Steps 1-4 are repeated for five different $\sigma_{rate}=0.1, 0.3, 0.5, 0.7, 0.9$ dex.
% \end{enumerate}

We adopt the following method to assess the impact of changing the thermal collision rates. Firstly, each thermal collision rate in the network (as seen in Table~\ref{table:1}) is given a log-normal probability distribution, with a mean equal to the literature value used in the standard network and with a standard deviation $\sigma_{rate}$ (units of dex).
Each rate coefficient is then randomly sampled from this distribution to make a randomly sampled network.
The CO mass is solved for using a full \textsc{Sumo} run at 300 days using the standard SN 1987A single-zone model for this randomly sampled network.  
These first steps are then are repeated 500 times, re-sampling the thermal collision rates at each instance to investigate the parameter space.
This investigation is are repeated for five different $\sigma_{rate}=0.1, 0.3, 0.5, 0.7, 0.9$ dex.

From these Monte Carlo simulations, we get a CO mass distribution for each $\sigma_{rate}$, which will indicate how the uncertainties of individual rates propagate to the molecular mass uncertainty.

\subsubsection{Results of MC simulations}

The left panel of Fig. \ref{fig:mc1} shows examples of the resulting CO mass distributions, for Monte Carlo runs with three different $\sigma_{rate}$. 
As seen, the distributions of CO mass can be fit well with a Gaussian in log-normal space. 
A larger $\sigma_{rate}$ leads to a larger variation in CO masses, as expected. 
Furthermore, the mean of the CO mass distribution shifts with $\sigma_{rate}$; with a larger $\sigma_{rate}$ the mean becomes larger.

As the resulting distribution of CO masses can be fitted with a Gaussian, we may compare the standard deviation of the CO mass distribution $\sigma_{CO}$ with the standard deviation of the individual rate coefficients $\sigma_{rate}$, seen in the right panel of Fig. \ref{fig:mc1}. 
For the values, $\sigma_{rate} = 0.1-0.9,$ there is a linear relationship between $\sigma_{rate}$ and $\sigma_{CO}$ with a slope slightly smaller than 1.
This indicates that there is a roughly linear relationship between the uncertainties of the thermal collision rate coefficients and the calculated mass of CO. 
Consequently, if the typical rate coefficient is known to a factor of two, then the CO mass uncertainty will be approximately a factor of two.

\section{Constraining physical parameters from molecule abundances}
\label{sect:constr}

   \begin{figure}
   \centering
   \includegraphics[width=\hsize]{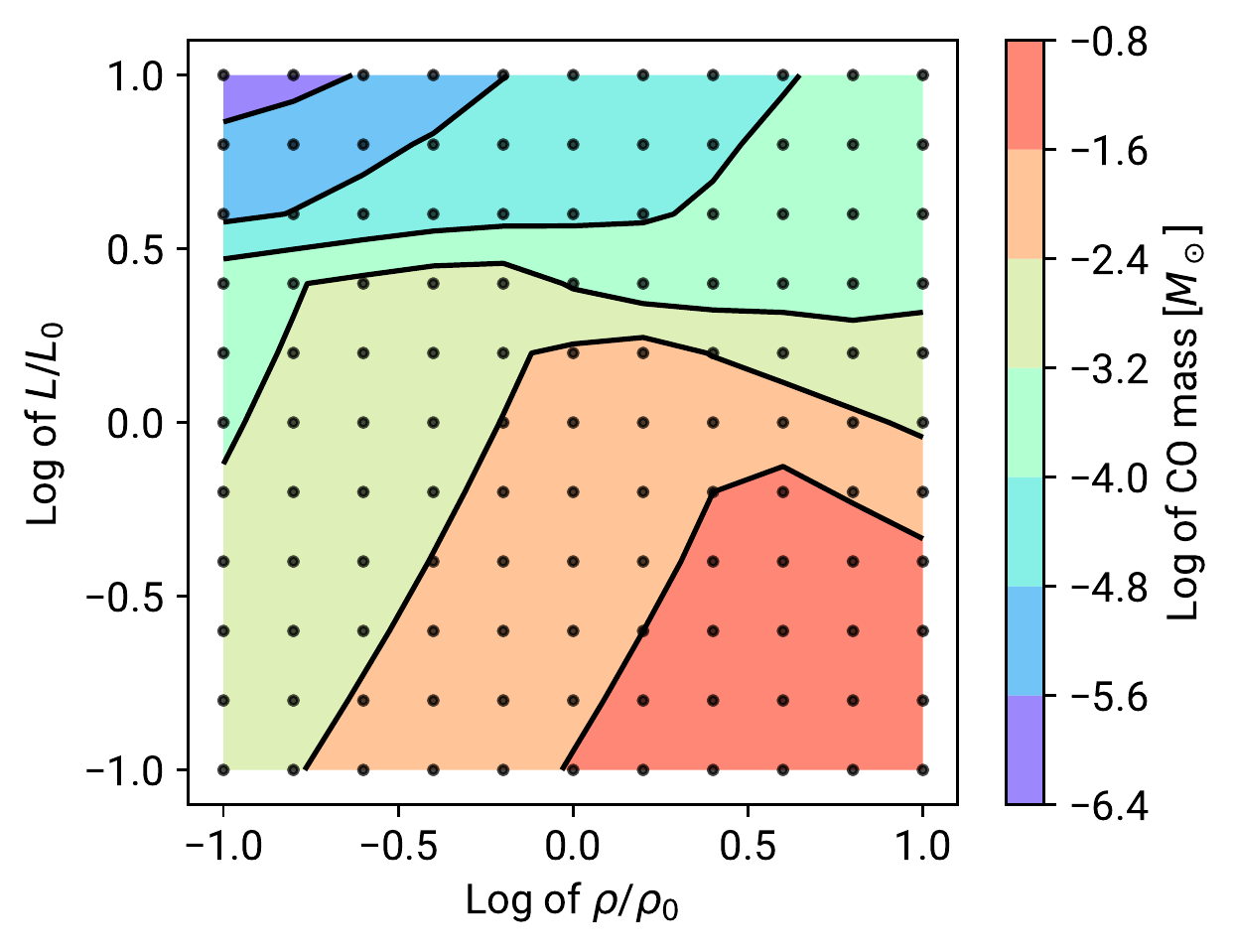}
      \caption{Variation of CO mass with deposition energy and density of the standard SN 1987A model at 284d. Each dot represents a model. }
         \label{fig:depdenco}
   \end{figure}

\begin{figure}
\centering
\includegraphics[width=0.24\textwidth]{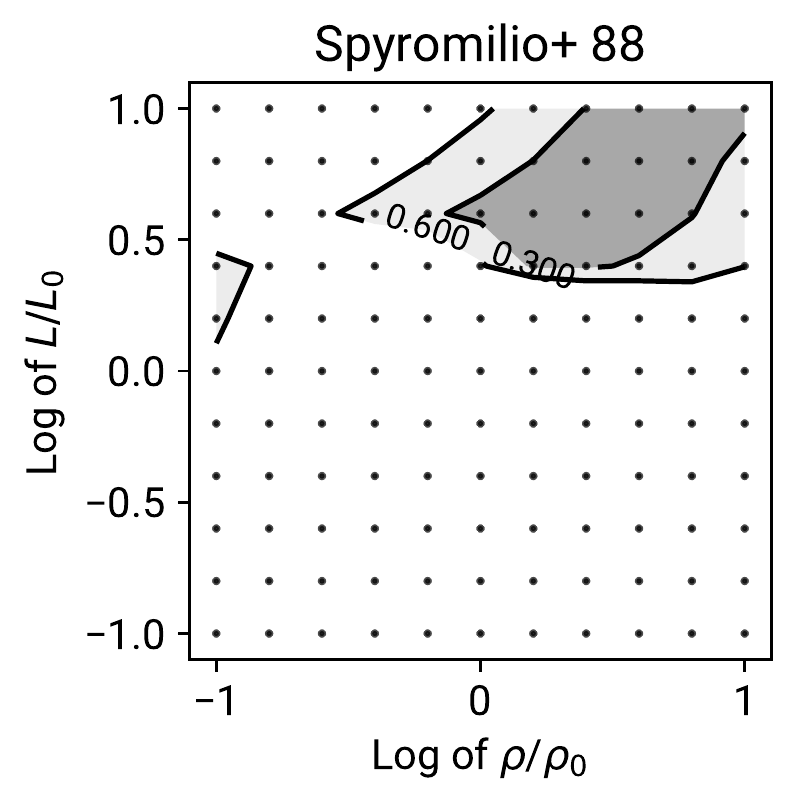} 
\includegraphics[width=0.24\textwidth]{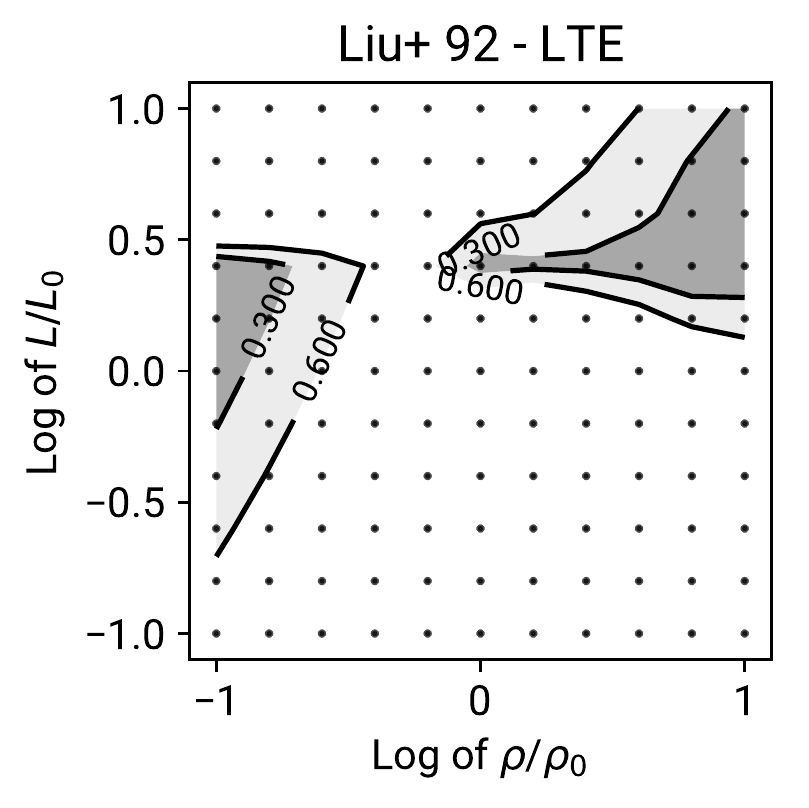}
\includegraphics[width=0.24\textwidth]{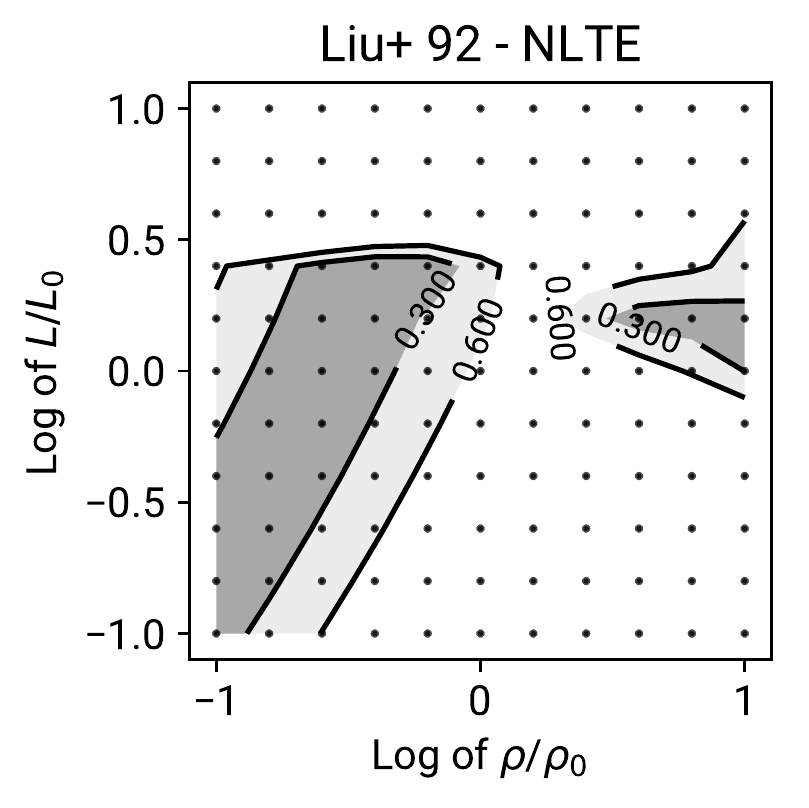}
\includegraphics[width=0.24\textwidth]{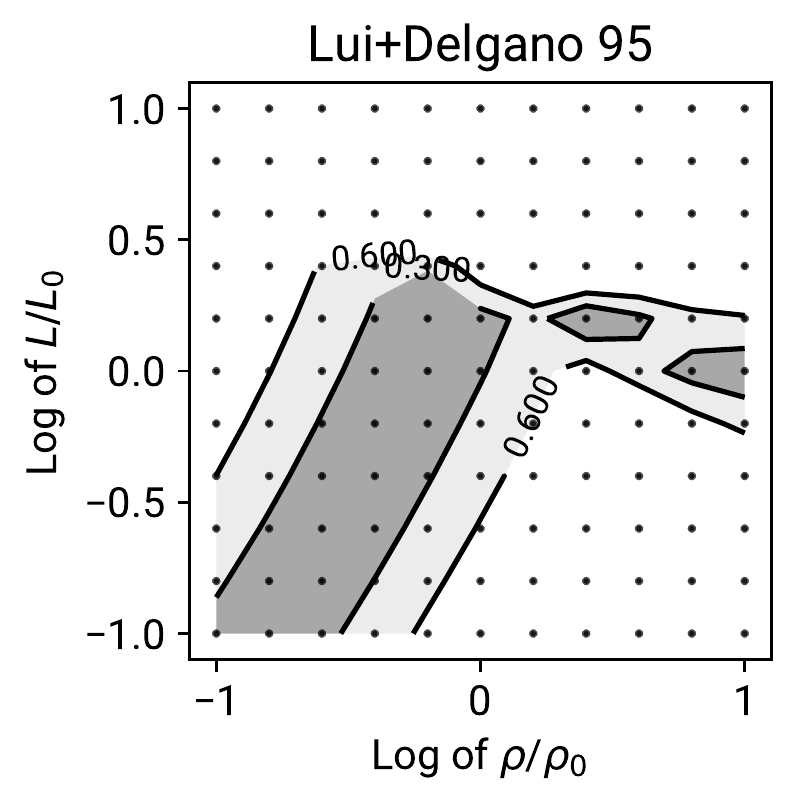}
\caption{Differences between various observational estimates and the model result, over a grid of models with varying deposition energies and densities. The upper two plots compare to LTE CO mass estimates from \citet{spyromilio_carbon_1988} and \citet{liu_carbon_1992}. The lower two plots are comparisons with NLTE CO estimates from \citet{liu_carbon_1992} and \citet{liu_oxygen_1995}. The dark grey shaded areas show the contours of 0.3 dex difference in CO mass and the light grey shaded areas show the 0.6 dex contours. $\rho_0=4.88 \times 10^{9}$cm$^{-3}$ , and $L_0=6.2 \times 10^{39}$erg s$^{-1}$  are the density and deposition energy for the standard SN 1987A model.} 
\label{fig:l1norm}
\end{figure}

If we can assume that the chemistry modeling parameters (i.e., abundance of C and O, the mass of the O/C zone, and the rate coefficients) are correct for a given SN within reasonable uncertainties, it is possible to use the modeled CO mass to constrain the deposition energy and the density by making a comparison with observational estimates of the CO mass. 
To investigate the feasibility of this method, we made a grid of models spanning a range of deposition energies and densities, with values ranging between $0.1 \times \rho_0$ and $10 \times  \rho_0$ for density, and $0.1 \times L_0$ and $10 \times L_0$ for deposition energy ($\rho_0=4.88 \times 10^{9}$cm$^{-3}$ , and $L_0=6.2 \times 10^{39}$erg s$^{-1}$  are the density and deposition energy for the standard SN 1987A model, described in Sect. \ref{Sect:res}, at 284 days.
The epoch of 284 days is chosen to coincide with the observations.
We produced 100 models, equidistant in log-space for the mentioned densities and deposition energies. 
For other input parameters, the same values were used as for the standard model (described in Sect.~\ref{sect:coform}).

The resulting CO mass for the model grid can be seen in Fig. \ref{fig:depdenco}. There is a clear inverse relationship between the deposition energy and CO mass, as discussed in Sect. \ref{sect:phys_par}. 
The density has a more complex relationship at these times, as the \ch{O^+} charge transfer reaction (Eq.~\ref{eq:oplus}) is still efficient for some densities.
As seen in Fig.~\ref{fig:co_res2}, 284 days is close to the transition point to the regime where increased density leads to higher CO mass, which is also the general trend seen in Fig.~\ref{fig:depdenco}.

The panels in Fig. \ref{fig:l1norm} show the norm between the log of CO masses from the model grid and from different observational estimates, (i.e., $\epsilon = |\text{log}(M_{CO}(\text{model})) - \text{log}(M_{CO}(\text{observation}))|$).
Consequently, the 0.3 contour delineates where the difference between the model and observed CO masses is 0.3 dex.
As shown in Fig. \ref{fig:depdenco}, bands through the $(L,\rho)$ plane produce the same CO mass. A density specification would, in general, allow a unique deposition energy to be picked out. However, a deposition energy specification can in some parts of the plane give degenerate density solutions. The more complex dependency on density is discussed in Sect. \ref{sect:phys_par}.
One particular conclusion can be drawn: if we compare to the NLTE CO mass estimates from \cite{liu_carbon_1992} and \cite{liu_oxygen_1995}, which generally produce the best fits to the spectral shapes, an upper limit on the deposition energy in the O/C zone of SN 1987A can be set of $\sim 2L_0 = 1.2\e{40}$ erg $^{-1}$, at 284 days. Such constraints can be used to test modern 3D hydrodynamic models where the energy deposition in any given zone depends on the morphology and mixing, which, in turn, depends on the properties of the explosion and the progenitor.

When more molecule mass estimates become available and can be modeled, as the correlation between deposition energy, density, and molecular mass depends on the unique creation and destruction pathways of each molecule, it could be possible to break degeneracies and get very specific constraints from observationally inferred molecule masses.

\section{Summary and conclusions}
\label{sect:sumcon}
In this work, we implement molecule formation and NLTE cooling into the spectral synthesis code \textsc{Sumo} and explore molecular formation physics as well as sensitivities to uncertain reaction rates. We study CO formation in a simplified model of the O/C zone of SN 1987A for a time interval of 150 to 600 days after the explosion. To summarize the results:
\begin{itemize}
    \item The thermal evolution is closely connected to the CO formation. The CO emission in the infrared adds a new significant cooling channel and after some time cooling by CO dominates in this environment. We find a positive feedback between CO formation and cooling. Consequently, self-consistent inclusion of CO cooling is important when simulating CO production.
    \item The CO mass uncertainty scales approximately linearly with the uncertainties of the thermal collision rates. If the rates are known to a factor of two, the CO mass uncertainty will be a factor two or slightly less (our formal relation between CO mass change and reaction rate uncertainty is 0.8).
    \item The physical parameters are important for CO production; density has a positive correlation with CO mass while deposition energy has a negative correlation. This could be used to constrain the physical parameters of the supernova if accurate CO mass observational estimates are available. Using this approach we can put an upper limit on the deposition energy at 284 days in the O/C zone of  SN 1987A to $\sim 2 \times L_0$, where $L_0$=$6.2 \times 10^{39}$erg s$^{-1}$ is the deposition energy of our standard model of SN 1987A.
    \item Our test model reproduces the observationally inferred CO mass in SN 1987A to a satisfactory degree. We have especially good agreement with the \citet{liu_oxygen_1995} estimate, for which NLTE and optical depth effects are taken into account (for comparison to other works see also Table \ref{table:workcomp}). 
    The CO mass settles at $\sim 4\e{-3}$ \msun~at 250d and stays there until at least 600d. This corresponds to a condensation efficiency of about 1\% of the carbon and oxygen. The CO lowers the temperature in the O/C zone by several 1000 degrees already from the beginning of the nebular phase; this means that the carbon and oxygen in this zone will not contribute significantly to optical emission.
\end{itemize}

It should be noted that the accuracy of the modeled CO mass hinges on both completeness of the network, meaning that most of the important and relevant reactions are accounted for, and on the accuracy of the used rate coefficients.
At times before 250 days, a single reaction, namely, the charge transfer between \ch{CO} and \ch{O+} (Eq.~\ref{eq:oplus}), dominates the destruction rate. 
If this reaction is disregarded, or if there are reactions of similar importance missing, the CO mass could potentially be quite different in that phase.
Similarly, the reported rate coefficients that are crucial for determining the molecular masses can vary greatly between different works and different methods. 
One example of this is the important radiative association reaction between C and O (\ch{C + O -> CO + h}$\nu$, Eq.~\ref{eq:raco}). 
In this work, the value from \citet{singh_radiative_1999} is used (RC37 in Table \ref{table:1}), which at 1500 K has a value of $1.77\times10^{-17}$ cm$^{3}$s$^{-1}$. 
This is almost a factor of two less than the value reported by \citet{gustafsson_radiative_2015} ($2.84\times10^{-17}$ cm$^{3}$s$^{-1}$). 
Other works report a range of values, for example, 
$1.08\times10^{-17}$ cm$^{3}$s$^{-1}$ (by 
\citealt{dalgarno_radiative_1990}, fit to data made by \citealt{cherchneff_chemistry_2009}), and 
$2.01\times10^{-17}$ cm$^{3}$s$^{-1}$ by \citet{franz_co_2011}.  
This is one of the most well-studied reaction rates; the situation for other reactions is more dire and the uncertainties of the reaction rates are an important source of uncertainty for the calculated molecular masses. 
Continued research into the reaction rates and different processes that are relevant to molecular formation in these environments is therefore of great importance for understanding supernova chemistry.

There are several aspects of the molecule treatment in \textsc{Sumo} that could be improved and made more realistic, with the end goal of obtaining more realistic synthetic SN spectra. 
A few planned improvements are mentioned here. 
While initially stated to only be a minor contribution to the destruction of CO, the photoionization of molecules should be implemented into the code.
As \textsc{Sumo} models the full radiation field of the ejecta, we can calculate the photoionization rates of molecules consistently, in contrast to previous works where these rates are estimated or inferred. 
Related to this is implementing self-consistent destruction rates of molecules by Compton electrons, which also potentially could be directly calculated in the code, instead of using previous estimates.
Finally, to produce realistic spectra, full multi-zone models are needed.
As these models will contain many more species, the chemical network used needs to be expanded to include the formation of more types of molecules. 
A possible time dependency should also be investigated for modeling at yet later phases.

\begin{acknowledgements}
The authors acknowledge funding from the Swedish National Space Board  (SNSB/Rymdstyrelsen), grant 2017-R 95/17. 
\end{acknowledgements}

\bibliographystyle{aa}
\bibliography{ref}

\begin{appendix}

\section{$W_i$ vs fractional ionization interpolation \label{polyfit_1}}
   \begin{figure}
   \centering
   \includegraphics[width=\hsize]{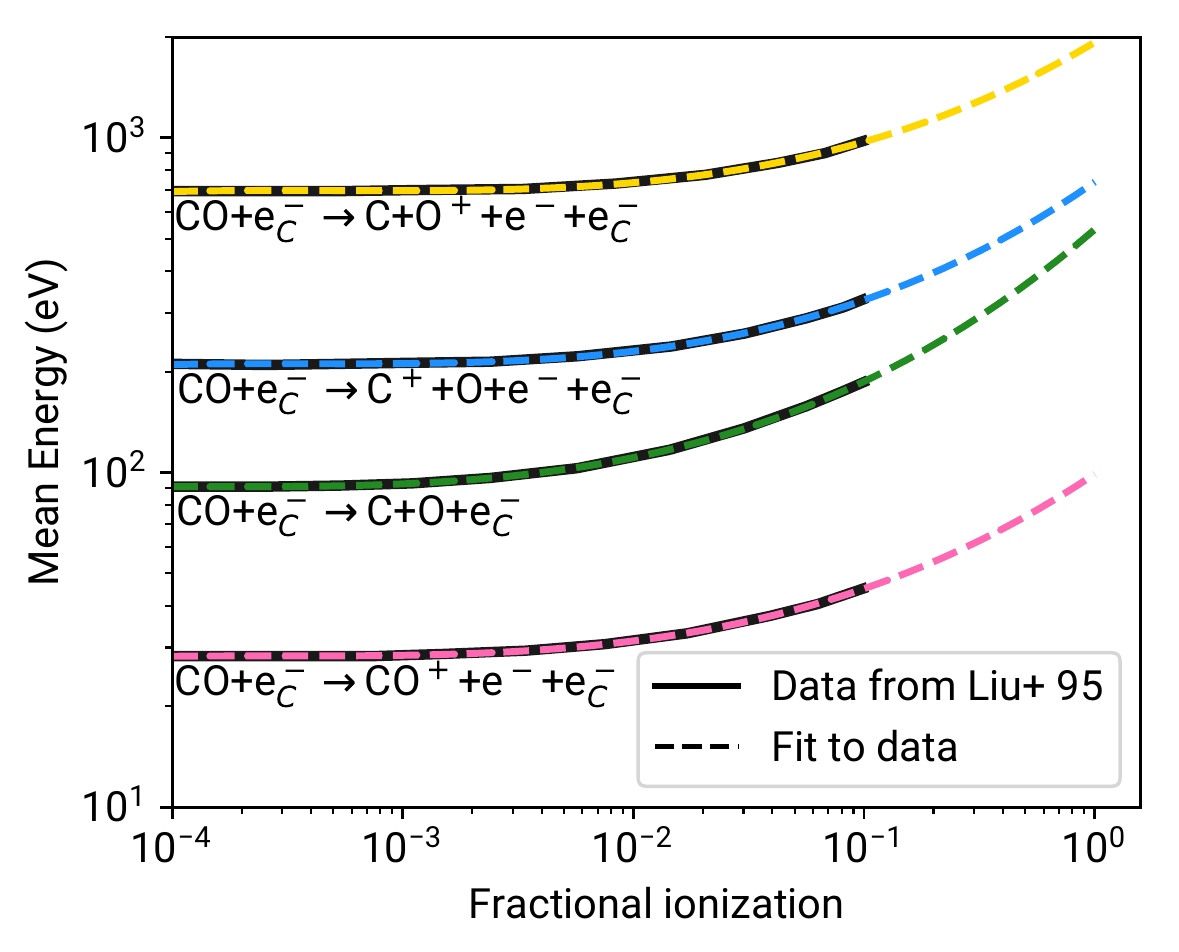}
      \caption{Original data and fits to the results on mean energy per ion pair for CO ionization presented in \citet{liu_oxygen_1995}. The original data is shown as solid lines and the fit to each relation is shown as dashed lines.}
         \label{fig:co_polyfit}
   \end{figure}

Figure 2. in \citet{liu_oxygen_1995} shows $W_i$, the mean energy per ionization or dissociation, as a function of the fractional ionization $x_e$, for different products of the reaction between CO and Compton electrons \ch{e^{-}_C}.
A cubic polynomial is fitted to this data in log-log space, to describe the variation of $W_i$ against $x_e$.
The fit takes the form 
 \begin{equation}
     \log(W_i) = a \times \log(x_e)^3 + b \times \log(x_e)^2 + c \times \log(x_e) + d
      \label{polyfit_2}
 \end{equation}
and can be seen in Fig. \ref{fig:co_polyfit}. 
The coefficients $a$, $b$, $c,$ and $d$ for each fit is shown in Table \ref{tab:fit}.
\begin{table}
\caption{Coefficients $a$, $b$, $c,$ and $d$, used in Eq.~\eqref{fig:co_polyfit}, for different products from the reaction \ch{CO + e_C^-}. }             % title of Table
\label{tab:fit}
\centering                          % used for centering table
\begin{tabular}{l l l l l}        % centered columns (4 columns)
\hline\hline                 % inserts double horizontal lines
Products & a & b & c & d \\    % table heading 
\hline                        % inserts single horizontal line
\ch{CO^+ + e^-}& 0.012& 0.13 & 0.46& 2.00 \\
\ch{C + O}& 0.013& 0.15 & 0.59& 2.73 \\
\ch{C^+ + O + e^-}& 0.014& 0.14 & 0.48& 2.87 \\
\ch{C + O^+ + e^-}& 0.013& 0.13 & 0.41& 3.29 \\
\hline                                   %inserts single line
\end{tabular}
\end{table}

\section{Model reaction rates}
\label{ap1}

  \begin{table*}
\caption{Overview of the reaction coefficients for thermal collision reactions. }
% \citet{mcelroy_umist_2013} is referred to as UMIST12, \citet{cherchneff_chemistry_2009} as C09, and \citet{sluder_molecular_2018} as S19.}             
\label{table:1}      
\centering       
\begin{tabular}{llrclrrrl}
\hline
 ID   & Reaction type         &            &     Reaction      &              &   $\alpha$ &   $\beta$ &   $\gamma$ & Ref.    \\
\hline
 RC1  & Charge Exchange       & \ch{C2 + O2+ } & \ch{->} &\ch{C2+  + O2}   &   4.1e-10  &      0    &     0   & 1$^a$ \\
 RC2  & Charge Exchange       & \ch{C2 + O+ } & \ch{->} &\ch{C2+  + O}     &   4.8e-10  &      0    &     0   & 1$^a$ \\
 RC3  & Charge Exchange       & \ch{C + C2+ } & \ch{->} &\ch{C+  + C2}     &   1.1e-10  &      0    &     0   & 1$^a$ \\
 RC4  & Charge Exchange       & \ch{C + CO+ } & \ch{->} &\ch{C+  + CO}     &   1.1e-10  &      0    &     0   & 2$^a$ \\
 RC5  & Charge Exchange       & \ch{C + O2+ } & \ch{->} &\ch{C+  + O2}     &   5.2e-11  &      0    &     0   & 1$^a$ \\
 RC6  & Charge Exchange       & \ch{C+  + C2O} & \ch{->} &\ch{C + C2O+ }   &   1e-09    &     -0.5  &     0   & 3$^a$ \\
 RC7  & Charge Exchange       & \ch{CO2+  + O2} & \ch{->} &\ch{CO2 + O2+ } &   5.3e-11  &      0    &     0   & 2$^a$ \\
 RC8  & Charge Exchange       & \ch{CO2+  + O} & \ch{->} &\ch{CO2 + O+ }   &   9.62e-11 &      0    &     0   & 4$^a$ \\
 RC9  & Charge Exchange       & \ch{CO + O+ } & \ch{->} &\ch{CO+  + O}     &   4.9e-12  &      0.5  &  4580   & 3$^a$ \\
 RC10 & Charge Exchange       & \ch{CO+  + C2} & \ch{->} &\ch{CO + C2+ }   &   8.4e-10  &      0    &     0   & 5$^a$ \\
 RC11 & Charge Exchange       & \ch{CO+  + CO2} & \ch{->} &\ch{CO + CO2+ } &   1e-09    &      0    &     0   & 1$^a$ \\
 RC12 & Charge Exchange       & \ch{CO+  + O2} & \ch{->} &\ch{CO + O2+ }   &   1.2e-10  &      0    &     0   & 1$^a$ \\
 RC13 & Charge Exchange       & \ch{CO+  + O} & \ch{->} &\ch{CO + O+ }     &   1.4e-10  &      0    &     0   & 2$^a$ \\
 RC14 & Charge Exchange       & \ch{O+  + O2} & \ch{->} &\ch{O + O2+ }     &   1.9e-11  &      0    &     0   & 6$^a$ \\
 RC15 & Ion-Neutral           & \ch{C2 + O2+ } & \ch{->} &\ch{CO + CO+ }   &   4.1e-10  &      0    &     0   & 7$^a$ \\
 RC16 & Ion-Neutral           & \ch{C2 + O+ } & \ch{->} &\ch{C + CO+ }     &   4.8e-10  &      0    &     0   & 7$^a$ \\
 RC17 & Ion-Neutral           & \ch{C2+  + O2} & \ch{->} &\ch{CO + CO+ }   &   8e-10    &      0    &     0   & 1$^a$ \\
 RC18 & Ion-Neutral           & \ch{C2+  + O} & \ch{->} &\ch{C + CO+ }     &   3.1e-10  &      0    &     0   & 3$^a$ \\
 RC19 & Ion-Neutral           & \ch{C + O2+ } & \ch{->} &\ch{CO+  + O}     &   5.2e-11  &      0    &     0   & 2$^a$ \\
 RC20 & Ion-Neutral           & \ch{C+  + CO2} & \ch{->} &\ch{CO + CO+ }   &   1.1e-09  &      0    &     0   & 8$^a$ \\
 RC21 & Ion-Neutral           & \ch{C+  + O2} & \ch{->} &\ch{CO + O+ }     &   4.54e-10 &      0    &     0   & 1$^a$ \\
 RC22 & Ion-Neutral           & \ch{C+  + O2} & \ch{->} &\ch{CO+  + O}     &   3-.42e-10 &      0    &     0   & 1$^a$ \\
 RC23 & Ion-Neutral           & \ch{CO2 + O+ } & \ch{->} &\ch{CO + O2+ }   &   9.4e-10  &      0    &     0   & 1$^a$ \\
 RC24 & Ion-Neutral           & \ch{CO2+  + O} & \ch{->} &\ch{CO + O2+ }   &   1.64e-10 &      0    &     0   & $^a$  \\
 RC25 & Neutral-Neutral       & \ch{C + CO2} & \ch{->} &\ch{CO + CO}       &   1e-15    &      0    &     0   & 9$^c$ \\
 RC26 & Neutral-Neutral       & \ch{C2 + O2} & \ch{->} &\ch{CO + CO}       &   1.5e-11  &      0    &  4300   & 10$^a$ \\
 RC27 & Neutral-Neutral       & \ch{C2 + O} & \ch{->} &\ch{C + CO}         &   2e-10    &     -0.12 &     0   & 10$^a$ \\
 RC28 & Neutral-Neutral       & \ch{C2O + O} & \ch{->} &\ch{CO + CO}       &   8.59e-11 &      0    &     0   & $^a$ \\
 RC29 & Neutral-Neutral       & \ch{C + C2O} & \ch{->} &\ch{CO + C2}       &   2e-10    &      0    &     0   & $^a$ \\
 RC30 & Neutral-Neutral       & \ch{C + CO} & \ch{->} &\ch{C2 + O}         &   2.94e-11 &      0.5  & 58025   & 11$^a$  \\
 RC31 & Neutral-Neutral       & \ch{C + O2} & \ch{->} &\ch{CO + O}         &   5.56e-11 &      0.41 &   -26.9 & $^a$ \\
 RC32 & Neutral-Neutral       & \ch{CO2 + O} & \ch{->} &\ch{CO + O2}       &   2.46e-11 &      0    & 26567   & $^a$ \\
%  RC33 & Neutral-Neutral       & \ch{CO + O2} & \ch{->} &\ch{CO2 + O}       &   5.99e-12 &      0    & 24075   & $^a$  \\
 RC33 & Neutral-Neutral       & \ch{CO + O2} & \ch{->} &\ch{CO2 + O}      &   4.2e-12  &      0    & 24053   & $^b$      \\
 RC34 & Neutral-Neutral       & \ch{CO + O} & \ch{->} &\ch{C + O2}         &   1e-16    &      0    &     0   & 12$^c$ \\
 RC35 & Radiative Association & \ch{C + C} & \ch{->} &\ch{C2}              &   4.36e-18 &      0.35 &   161.3 & 13$^a$ \\
 RC36 & Radiative Association & \ch{C + C+ } & \ch{->} &\ch{C2+ }          &   4.01e-18 &      0.17 &   101.5 & 13$^a$ \\
 RC37 & Radiative Association & \ch{C + O} & \ch{->} &\ch{CO}              &   4.69e-19 &      1.52 &   -50.5 & 14$^a$ \\
 RC38 & Radiative Association & \ch{C + O+ } & \ch{->} &\ch{CO+ }          &   5e-10    &     -3.7  &   800   & 4$^a$ \\
 RC39 & Radiative Association & \ch{C+  + O} & \ch{->} &\ch{CO+ }          &   3.14e-18 &     -0.15 &    68   & 1$^a$ \\
 RC40 & Radiative Association & \ch{O + O} & \ch{->} &\ch{O2}              &   4.9e-20  &      1.58 &     0   & 1$^a$ \\
\hline
\end{tabular}
\tablefoot{When possible the original works are referenced.\\
$^a$ Obtained from the UMIST database (\citealt{mcelroy_umist_2013},  \href{http://udfa.ajmarkwick.net/}{www.astrochemistry.net}). \\
$^b$ Obtained from \citet{cherchneff_chemistry_2009} \\
$^c$ Obtained from \citet{sluder_molecular_2018}.}

\tablebib{
(1)~\citet{prasad_model_1980};
(2) \citet{adams_reactions_1978}; 
(3) \citet{fehsenfeld_thermal_1972}; 
(4) \citet{petuchowski_co_1989}; 
(5) \citet{copp_selected_1982}; 
(6) \citet{fahey_rate_1981}; 
(7) \citet{oscar_martinez_gas_2008}; 
(8) \citet{viggiano_laboratory_1980};
(9) \citet{mitchell_effects_1984};
(10) \citet{smith_rapid_2004}; 
(11) \citet{wakelam_kinetic_2012};  
(12) \citet{dalgarno_radiative_1990};
(13) \citet{andreazza_formation_1997};
(14) \citet{singh_radiative_1999}.
}
\end{table*}
 
\begin{table*}
\caption{Overview of the coefficients for the molecular recombination reactions.}             
\label{table:2}      
\centering       
  
  \begin{tabular}{llrclrrrl}
\hline
 ID   & Reaction type         & &Reaction&                                    &   $\alpha$ &   $\beta$ &   E$_a$ & Ref.    \\
\hline
 RR1  & Dissociate recombination       & \ch{CO+ + e- } & \ch{->} & \ch{C + O + h$\nu$}   &   2.36e-12  &      -0.29   &     -17.6   & 1\tablefootmark{a} \\
 RR2  & Dissociate recombination       & \ch{CO2+ + e- }& \ch{ ->} & \ch{CO + O + h$\nu$}   &   3.8e-07  &      -0.5    &     0   & 1\tablefootmark{a} \\
 RR3  & Dissociate recombination       & \ch{C2+ + e-  }& \ch{->} & \ch{C + C + h$\nu$}   &   3e-07  &      -0.5    &     0   & 1\tablefootmark{a} \\
 RR4  & Dissociate recombination       & \ch{C2O+ + e- }& \ch{ ->} & \ch{CO + O + h$\nu$}   &   3e-07  &      -0.5    &     0   & \tablefootmark{a} \\
 RR5  & Dissociate recombination       & \ch{O2+ + e- } & \ch{->} & \ch{O + O + h$\nu$}   &   1.95e-07 & -0.7    &     0   & 2\tablefootmark{a}\\

\hline
\end{tabular}
\tablefoot{When possible the original works are referenced.\\
$^a$ Obtained from the UMIST database (\citealt{mcelroy_umist_2013}, \href{http://udfa.ajmarkwick.net/}{www.astrochemistry.net}).}
\tablebib{(1)~\citet{brian_dissociative_1990};
(2) \citet{alge_measurements_1983}
}

\end{table*}

The reaction rates used in the SN 1987A test model. 
As this model only contains carbon and oxygen, and ions of carbon and oxygen; all included molecules are composites of these. 

\section{Deposition energy \label{ap:depe}}

\begin{figure}
   \centering
   \includegraphics[width=\hsize]{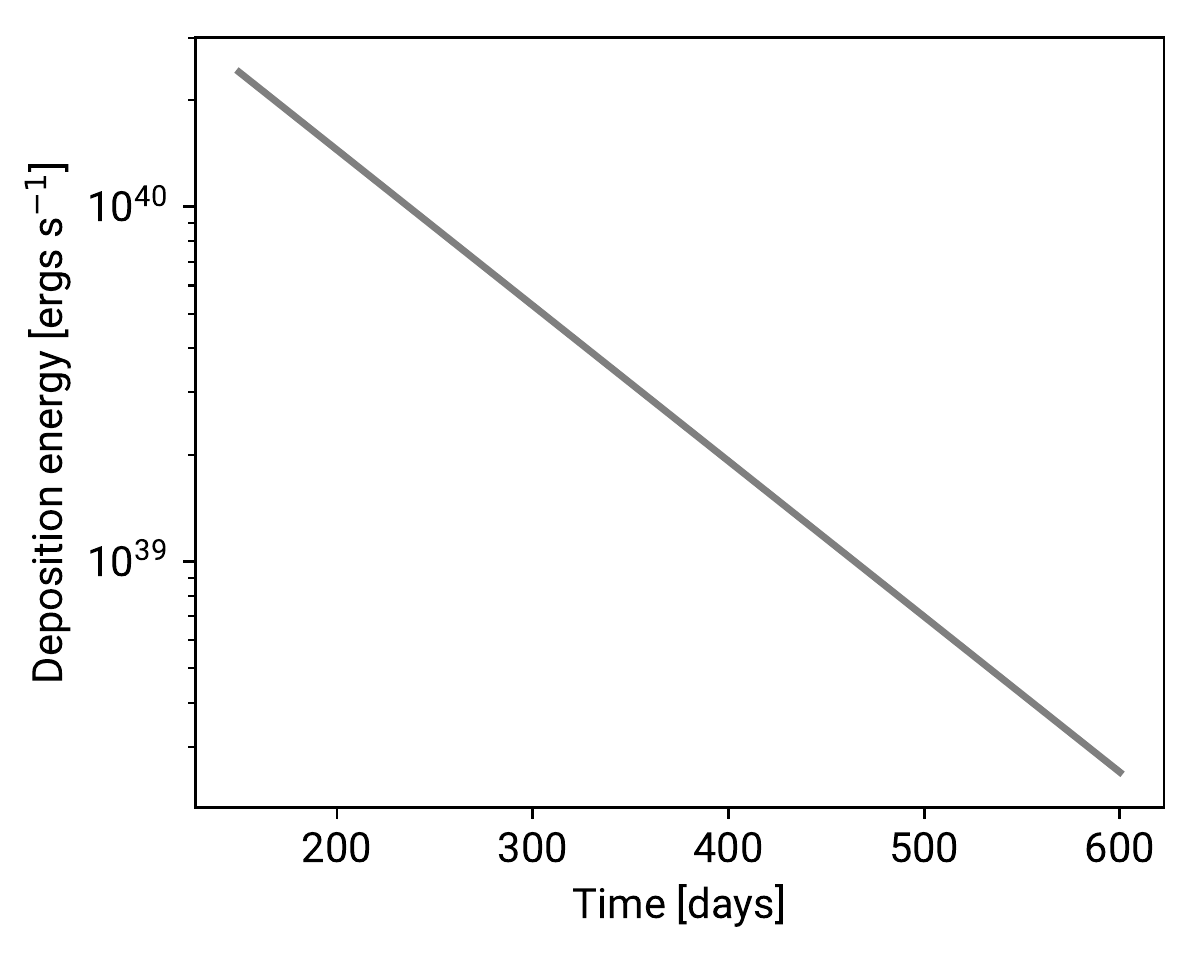}
      \caption{Deposition energy used for the standard SN 1987A model, taken from the calculations of \citet{jerkstrand_progenitor_2012}.}
         \label{fig:depe}
   \end{figure}

Fig. \ref{fig:depe} shows the deposition energy used for the standard SN 1987A model, which is adopted from \citet{jerkstrand_progenitor_2012}. 

\section{Overview of other work \label{ap:workcomp}}
Overview of the physical parameters in models for CO formation SN 1987A, from different works.
\begin{sidewaystable*}
\caption{Comparison of key properties of different chemical models describing CO formation in SN 1987A. }
\label{table:workcomp}
\centering
\begin{tabular}{l|l|l|l|l|l|l}

\hline\hline  
Work & CO mass & Physical model & Composition & C/O & Number density [cm$^{-3}$]& Temperature [K] \\
\hline  \hline  

\citet{petuchowski_co_1989} 
& $\sim 2 \times 10^{-5}$M$_\odot^a$
& \begin{tabular}[c]{@{}l@{}}Model adapted\\from \citet{woosley_sn_1988}.\end{tabular} 
& \begin{tabular}[c]{@{}l@{}} 1.25M$_\odot$ O, 0.2M$_\odot$ C,...,\\ (see Table. 1 in \\ \citealt{petuchowski_co_1989})\end{tabular} 
&  0.21
& $1.3 \times 10^{9}$ 
& $4000$ \\

\hline  

 \citet{lepp_molecules_1990}& \begin{tabular}[c]{@{}l@{}}$5 \times 10^{-7}$M$_\odot^b$, \\ $2 \times 10^{-6}$M$_\odot^b$,\\ $5 \times 10^{-5}$M$_\odot^b$,\\ (284 days)\end{tabular} 
 &  \begin{tabular}[c]{@{}l@{}}10HMM model from\\ \citet{pinto_theory_1988}\end{tabular} 
 & \begin{tabular}[c]{@{}l@{}}1.1M$_\odot$ O, 0.16M$_\odot$ C, ...,\\ (see text in
 \citealt{lepp_molecules_1990})\end{tabular} 
 &  0.19
 & $5.6 \times 10^{9}$ 
 & $1600$ (284 days) \\
 
 \hline  

\citet{liu_carbon_1992} 
& \begin{tabular}[c]{@{}l@{}}$\sim 5 \times 10^{-6}$M$_\odot^{a,c}$, \\ $\sim 2 \times 10^{-2}$M$_\odot^{a,c}$, \\ $\sim 2 \times 10^{-3}$M$_\odot^{a,c}$\end{tabular} 
& \begin{tabular}[c]{@{}l@{}}Mixed model from \\ \citet{nomoto_confronting_1991}. \end{tabular} 
& \begin{tabular}[c]{@{}l@{}}1.48M$_\odot$ O, 0.11M$_\odot$ C, ...,\\ (see Table 1 in \\ \citealt{nomoto_confronting_1991})\end{tabular} 
&  0.1
& $5.2 \times 10^{9}$ 
& $2500$ (284 days) \\

 \hline  

 \citet{liu_oxygen_1995}& $\sim 3 \times 10^{-3}$M$_\odot$ 
 & \begin{tabular}[c]{@{}l@{}}Same as \\\citet{liu_oxygen_1995} \end{tabular} 
 & \ditto
 & \ditto
 & $2.6 \times 10^{9}$ 
 & $\sim 2000^a$ \\
 
 \hline  
 
 \citet{sarangi_chemically_2013}& $\sim 8 \times 10^{-3}$M$_\odot^a$ 
 & \begin{tabular}[c]{@{}l@{}}15M$_\odot$ ZAMS model,\\ from  \citet{rauscher_nucleosynthesis_2002}$^d$\end{tabular} 
 &  \begin{tabular}[c]{@{}l@{}}4A$^e$: 0.15\msun O, 0.04 \msun C,...,\\4B$^e$: 0.11\msun O, 0.06\msun C ,.., \\(see Table 1 in \\\citealt{sarangi_chemically_2013})\end{tabular}
 &  \begin{tabular}[c]{@{}l@{}}4A: 0.37\\4B: 0.74\end{tabular}
 & \begin{tabular}[c]{@{}l@{}}4A: $1.6 \times 10^{10}$\\4B: $2.3 \times 10^{10}$\end{tabular} 
 & \begin{tabular}[c]{@{}l@{}}4A: 1998\\4B: 1899\end{tabular} \\
 
 \hline  
 
 \citet{sluder_molecular_2018}& $\sim 10^{-2}$M$_\odot^a$ 
 & \begin{tabular}[c]{@{}l@{}}20M$_\odot$ ZAMS model, \\ calculated using  \\the MESA code.\end{tabular} 
 & - 
 & -
 & - 
 & $\sim 1500^a$ \\
 
 \hline  
 
 This work& $4 \times 10^{-3}$M$_\odot$ 
 &  \begin{tabular}[c]{@{}l@{}}O/C zone from \\ \citet{woosley_nucleosynthesis_2007} \end{tabular}
 &  0.4\msun O, 0.18\msun C
 &  0.6
 &  $4.1 \times 10^{9}$
 & 1664 \\
 
 \hline  

\end{tabular}
\tablefoot{The presented values are obtained from the specified work (to the authors best knowledge) at 300 days unless specified otherwise, if the data was available.\\
$^a$ Extracted from plot. \\
$^b$ Results from three models; standard, and two different He destruction rates, respectively. \\
$^c$ Result from mixed model, unmixed model, and a semi-mixed model made to fit observations. \\
$^d$ Several models are investigated in this work, however the 15\msun model is shown here as this is the most in-depth discussed. \\
$^e$ This model have two zones significantly contributing to CO production; 4A - O/C zone and 4B - He/O/C zone. }
\end{sidewaystable*}

\end{appendix}

\end{document}